\documentclass[preprint,12pt]{elsarticle}
\usepackage{amssymb,amsmath,verbatim,bm,color,mathtools,multicol,relsize}
\usepackage{ulem}
\usepackage{bbm}
\usepackage{graphicx,tikz}
\usepackage{psfrag}
\usepackage{xcolor}
\usepackage{siunitx}
\usepackage{tabularx}
\usepackage{multirow,array,booktabs}
\usepackage{caption}
\usepackage{subcaption}
\usepackage{wrapfig}
\usepackage{upgreek}
\usepackage{float}

\newcommand{\sumnolim}{\sum\nolimits}
\newcommand{\smallsum}{\mathop{\mathsmaller{\sum}}\nolimits}

\newcommand{\expect}[1]{\mathrm{E}\left\{#1\right\}}
\newcommand{\Matlab}{$\mathrm{Matlab}$}
\newcommand{\matrixstyle}[1]{\mathrm{#1}}
\newcommand{\vectorstyle}[1]{\boldsymbol{\mathrm{#1}}}

\newcommand{\operatorstyle}[1]{\mathcal{#1}}

\journal{Mechatronics}

\begin{document}
	
\begin{frontmatter}

	\title{Inverse Parametric Uncertainty Identification using Polynomial Chaos and high-order Moment Matching benchmarked on a Wet Friction Clutch}
	
	\author{Wannes De Groote*\thanks{*Both authors contributed equally to this work.}\textsuperscript{1,2}, Tom Lefebvre*\textsuperscript{1,2}, Georges Tod\textsuperscript{3}, Nele De Geeter\textsuperscript{1,2}, Bruno Depraetere\textsuperscript{3}, \\ Suzanne Van Poppel\textsuperscript{4}, Guillaume Crevecoeur\textsuperscript{1,2}\\
		\textsuperscript{1} EEMMeCS, Ghent University, Belgium\\
		\textsuperscript{2} EEDT Decision \& Control, Flanders Make, Belgium \\
		\textsuperscript{3} DecisionS, Flanders Make, Belgium \\
		\textsuperscript{4} CodesignS, Flanders Make, Belgium
		\vspace*{-.75cm}}

	%
	%
	%
\begin{abstract}
	A numerically efficient inverse method for parametric model uncertainty identification using maximum likelihood estimation is presented. The goal is to identify a probability model for a fixed number of model parameters based on a set of experiments. To perform maximum likelihood estimation, the output probability density function is required. Forward propagation of input uncertainty is established combining Polynomial Chaos and moment matching. High-order moments of the output distribution are estimated using the generalized Polynomial Chaos framework. Next, a maximum entropy parametric distribution is matched with the estimated moments. This method is numerically very attractive due to reduced forward sampling and deterministic nature of the propagation strategy. The methodology is applied on a wet clutch system for which certain model variables are considered as stochastic. The number of required model simulations to achieve the same accuracy as the brute force methodologies is decreased by one order of magnitude.
	The probability model identified with the high order estimates resulted into a true log-likelihood increase of about 4\% since the accuracy of the estimated output probability density function could be improved up to 47\%. 
\end{abstract}


\end{frontmatter}


\section{Introduction}
The increasing performance demands in design and control of mechatronic applications lead to an upward trend in the need for accurate and robust models \cite{wilamowski2018control}. These mathematical relations allow to make predictions on the behavior and performance of the system in a virtual computational environment. It enables efficient design processes, leading to faster successive optimization iterations, so that more reliable products can be made at lower production cost. 

In this research, we consider the modeling of a wet clutch system. Wet friction clutches are hydraulic-mechanical devices used to transmit torque from an input shaft to an output shaft by means of friction. These systems are used in various types of automatic transmissions to selectively engage gear elements. During engagement, multiple plates make contact which enables power transfer from motor to load. Contrary to dry clutches, the plates are bathed in oil to assure better heat conduction of the friction losses. Wet friction clutches thus result in higher torques which make them suitable and primordial in off-road vehicles and agricultural machines \cite{Widanage2011}. Models of clutch systems have been widely used for fault diagnosis purposes \cite{foulard2015} and numerical optimization of both design and control \cite{della2018}. Clutch models identified for the wet clutch considered in this paper have been previously implemented in condition monitoring \cite{agusmian2013} and control \cite{bruno2011}. Additionally, model based control techniques applied on the wet clutch setup, have proven shorter convergence time and can avoid excessive control inputs that lead to unsafe operations \cite{dekeyser2014}.

\begin{figure}[t!]
	\centering
	\includegraphics[width=.9\columnwidth]{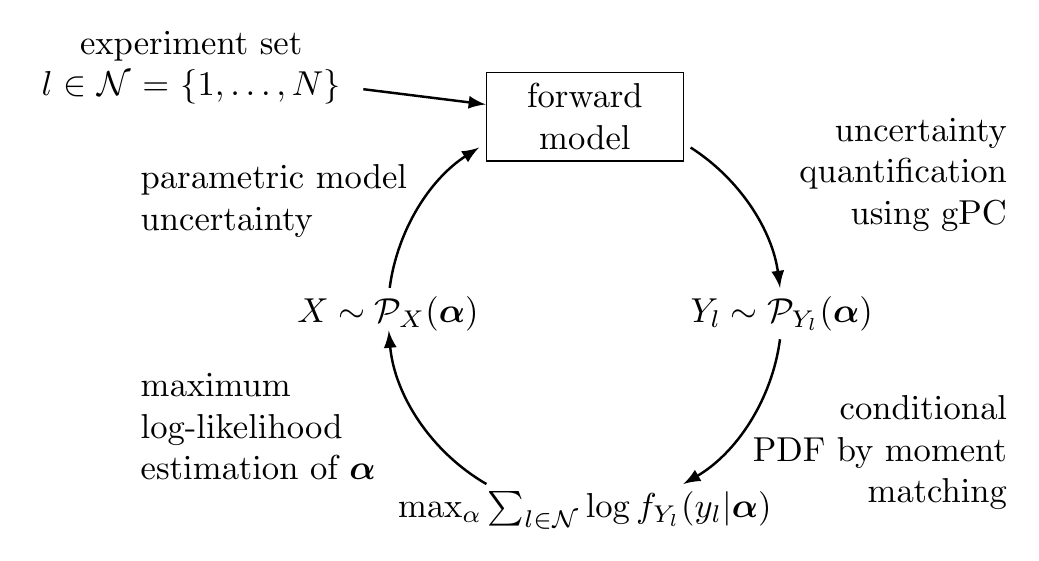}
	\caption{Proposed identification method determines optimal $\vectorstyle{\alpha}$, associated to $\mathcal{P}_X(\vectorstyle{\alpha})$ of the uncertain parameters within the model, via MLE. The input distribution $\mathcal{P}_X(\vectorstyle{\alpha})$ is propagated through the model towards a corresponding output distribution $\mathcal{P}_{Y_l}(\vectorstyle{\alpha})$ for each performed experiment $l \in \mathcal{N}$. The novelty lies in the sequential use of gPC and PDF construction based on moment matching of gPC based high-order moments.}
	\label{fig:approach}
\end{figure}
Although the value of having accurate and robust models for mechatronic applications, as illustrated with the wet friction clutch, is straightforward, the construction of accurate models remains a cumbersome process. Since these systems are plagued by their intrinsic complexity and nonlinear  behavior \cite{Widanage2011}, variations in the physical phenomena over time due to e.g. wear of plates or degradation and centrifugal effects of the oil in the wet friction clutches; model discrepancies are inevitable  \cite{watson2005}. Due to these uncertaintes, deterministic models become less useful. Alternatively, one can consider a system model of stochastic nature that predicts an output distribution for given control input. Models with ingrained parametric uncertainty have proven increased robustness in the field of fault diagnosis \cite{touati2012} and control \cite{marconi2008,abdeetedal2018}. In this research we adopt the available system model and focus on identifying a probabilistic model to several lumped model parameters. Mostly (mechatronic) models are derived from first principles. Lumped parameter values are then determined empirically by fitting the model to the experiments. Considering that this strategy generates a model that can explain the experiment set, it is no direct proof of the validity of the physical model. It may occur that the parameter value fitted on this experiment set compensates for an inherent model shortcoming. Rather than increasing the model complexity, we propose to couple a parameterized input probability distribution to some of the lumped model parameters \cite{hajiloo2012}. We assume that the uncertainty structure (i.e. the parameterized input probability distribution) is fixed, and only its parameters, such as mean and standard deviation (e.g. for a normal distribution), need to be identified based on what parameters best explain a number of experiments.

Stochastic system identification can be performed by means of prevailing methods as (quasi) Monte Carlo (MC) sampling techniques for assessing the propagation of the uncertainty parameter to the output variables (response) of the system model \cite{stein1987large,caflisch1998monte}. These techniques are however curtailed by the computational inconvenience that comes with the numerous forward simulations required to achieve an acceptable degree of accuracy. Furthermore, in an inverse uncertainty identification setting, the uncertain parameters need to be optimized with respect to the correspondence of model responses to measurements, leading to even higher computational costs. More recent work has revived the exploitation of generalized Polynomial Chaos (gPC) expansions that can lead to significant gains in terms of computational feasibility \cite{xiu2002wiener,xiu2003modeling,xiu2005high,xiu2007efficient,blatman2011adaptive}. In the gPC framework, the propagation of input uncertainty to output variables is realized by developing the stochastic subspace through a polynomial series expansion. We propose in this paper to choose the polynomial basis so that it is orthogonal with respect to the joint probability density function of the random input variables. Hence, an efficient mathematical context emerges that allows to express the statistical moments as a function of the polynomial coefficients. Once these moments are available, the conditional Probability Density Function (PDF) of the outcome of the experiment can be retrieved through the well-known method of moments \cite{bowman2004estimation,munkhammar2017polynomial}. 

To our knowledge, application of the gPC framework to estimate probabilistic moments of a stochastic output model has remained limited to mean and variance estimates. In this work we devised an efficient algorithm to calculate high-order moments as well, so to increase the information content that can be passed to the moment matching algorithm. The major computational bottleneck is thereby isolated and can be executed as an offline step. Figure \ref{fig:approach}  illustrates the incorporation of the identified output distributions for each experiment within a Maximum Likelihood Estimation (MLE) framework \cite{bickel2015}. This methodology enables to find sufficient parametric uncertainty that a model needs to include so that different experiments can be explained \cite{Berx2014}. A Genetic Algorithm (GA) is used here to optimize the parameterized distributions of the model parameters, because of its exploratory characteristics \cite{chan2007}. First, we elaborate our inverse uncertainty identification method to define different methods and concepts that are included within our theoretical framework. Next we address the wet clutch system and discuss in detail our practical set-up. Lastly, the developed methods are applied on the clutch application to identify the input probability model characterized by high-order moments. \\

\section{Inverse uncertainty identification method}
\label{sec:IUIM}
\subsection{Introduction}
We consider physics based nonlinear forward models, $\mathcal{Y}:\mathcal{N}\times\mathbb{R}^n \rightarrow\mathbb{R}$, used to generate predictions, $Y_l=\mathcal{Y}(l;\vectorstyle{x})$, for the univariate outcome of experiments, $y_l$, that are characterized by some index $l\in\mathcal{N}$. The index $l$ is assumed to contain sufficient information to render the experiments deterministic from the perspective of the model, i.e. the information contains the operational settings that can be adjusted by an experimenter to the mechatronic system. The variable  $\vectorstyle{x}\in\mathbb{R}^n$ contains any (lumped) circumstantial parameters that, once identified, are assumed to remain constant for all experiments.  
We consider a number ($N$) of experiments within the index set $\mathcal{N}=\{1,\cdots,N\}$. When we assume that the measurements are deterministic, the mismatch stems either from a lack of circumstantial knowledge or suggests a genuine shortcoming of the model. In the former case, a latent variable may be present while in the latter case, the model does not, or not correctly, take into account all underlying physical phenomena.\\
In recent literature issues are addressed by associating a random variable, $\vectorstyle{X}\in\mathcal{X}\sim\mathcal{P}_X$, to the model parameter set $\vectorstyle{x}$ \cite{hajiloo2012}. This approach renders the forward model stochastic and rather than exact predictions the model generates a distribution of possible outcomes. Although this strategy may deny an underlying physical reality, it grants the model practical use as it hands the user a measure for how far off a prediction may be. With respect to this concept, our contribution is twofold. \\
First, we propose an efficient numerical method to obtain an approximation of the nonlinear transformed output distribution of the random variable $Y_l = \mathcal{Y}(l;\vectorstyle{X})$, by moment matching with a parametric distribution. The more moments are taken into account, the better the approximation will be. To facilitate cheap yet reliable moment estimates we engage the gPC framework and extend it so that high-order moments can be extracted rigorously. Such benefits both the MLE, as nonlinear effects can be taken into account, as well as the post identification usage of the stochastic model.\\
Secondly, we propose a fast computational method to associate a probability model to the variable, $\vectorstyle{X}$, based on a number of experimental observations, $\{y_l\}_{l\in\mathcal{N}}$. To that end we put forward a parametric model, $\mathcal{P}_X(\vectorstyle{\alpha})$, for the probability of the random variable, $\vectorstyle{X}$, parameterized by the variable $\vectorstyle{\alpha}$, and we identify an optimal value for $\vectorstyle{\alpha}$ by means of Maximum Likelihood Estimation (MLE). Hereinafter, for notational convenience we make no distinction between model  $\mathcal{P}_X(\vectorstyle{\alpha})$ and parameter $\vectorstyle{\alpha}$, and refer to it as the (input) probability model.

\vspace*{-10pt}
\subsection{Maximum Log-Likelihood Estimate} The proposed identification procedure is formulated as an inverse problem (\ref{eq:MLE}). That is, we want to find the probability model, $\vectorstyle{\alpha}^*$, that maximizes the natural $\log$ of the likelihood, $ L(\vectorstyle{\alpha})$. The likelihood of a given probability model, $\vectorstyle{\alpha}$, is calculated as the product (or summation of the natural logarithm) of the probability density functions, $f_{Y_l}(y_l|\vectorstyle{\alpha})$ $l\in \mathcal{N}$. Probability $f_{Y_l}(y_l|\vectorstyle{\alpha})$ is the conditional relative likelihood that the value of the stochastic model output, $Y_l$, of experiments $l$, as obtained through simulation given probability model $\vectorstyle{\alpha}$, would equal the experimentally observed value, $y_l$. The higher the value of $L$, the more likely the forward model.

\begin{equation}
\begin{aligned}
\vectorstyle{\alpha}^* &= \max_{\vectorstyle{\alpha}} \log L\left(\vectorstyle{\alpha}\right)=\max_{\vectorstyle{\alpha}} \sum_{l\in\mathcal{N}} \log L_l \\ 
&= \max_{\vectorstyle{\alpha}} \sum_{l\in\mathcal{N}} \log f_{Y_l}\left(y_l|\vectorstyle{\alpha}\right)
\end{aligned}
\label{eq:MLE}
\end{equation}

With regard to the construction of the MLE cost function, $\log L$, it is clear that the computational bottleneck is predominated by the numerical evaluation of the conditional probabilities, $f_{Y_l}(y_l|\vectorstyle{\alpha}),l\in\mathcal{N}$. Hence, we will require an efficient numerical procedure to quantify the probabilities, $f_{Y_l}$. 

A novel procedure of such kind is described here next.

\vspace*{-10pt}
\subsection{Moment Estimation}
\label{sec:UP-PCE}
To evaluate the conditional output probability density functions, $f_{Y_l}\left(y_l|\vectorstyle{\alpha}\right)$ we will engage the generalized Polynomial Chaos (gPC) expansion framework. The gPC framework accounts for a number of advantageous mathematical conditions to propagate uncertainty. Such are accomplished by approximating the forward nonlinear model with a polynomial expansion. By subsequently choosing the polynomial basis so that it satisfies orthogonality conditions with respect to the probability density function of the random input variables, an efficient mathematical context emerges that allows to express the statistical moments in function of the coefficients associated to the expansion. Conventional application is limited to the mean and variance. In this work we propose a novel method to incorporate high-order moments as well.

\subsubsection{Generalized polynomial chaos}
According to the polynomial approximation theorem, any smooth function, $y:\mathbb{R}^n\rightarrow\mathbb{R}$, is equivalent to an infinite polynomial series \cite{ghanem1991stochastic}
\begin{equation*}
y(\vectorstyle{x}) = \sum\nolimits_{i=1}^{\infty} c_i \psi_i(\vectorstyle{x})
\end{equation*}

From a computational perspective our interest is however reserved to the $d$-th order approximation. That is, let $\mathcal{P}^d_n$ be the $n$-variate polynomial space of at most degree $d$ and let $\vectorstyle{\psi} = \{\psi_i\}_{i=1}^p$ serve as basis for $\mathcal{P}^d_n$ with $p = \frac{(n+d)!}{n!d!}$. The $d$-th order approximation is then given by (\ref{eq:poly-D}) for given polynomial coefficients $\{c_i\}_{i=1}^p$. An $n$-variate basis $\vectorstyle{\psi}$ can be constructed from $n$ univariate bases $\vectorstyle{\phi}^{(k)} = \{\phi^{(k)}_j\}_{j=0}^d, k\in\{1,\dots,n\}$. Consider the basis vector elements, $\psi_{|\uline{i}|\leq d} = \prod_{k=1}^{n} \phi^{(k)}_{\uline{i}(k)}$
, with multi-index $\uline{i} = (i_1,\dots,i_n)$ and where $|\uline{i}|$ is defined as $\sum_{k=1}^{n}i_k$. For notational convenience we exploit the bijection between index $\uline{i}$ and $i$ taking values in $\mathcal{I}=\{1,\dots,p\}$.
\begin{equation}
\label{eq:poly-D}
y^{(d)}(\vectorstyle{x}) = \sum\nolimits_{i\in\mathcal{I}} c_i \psi_i(\vectorstyle{x})
\end{equation}

Within the context of uncertainty propagation this representation allows to establish advantageous computational conditions by a distinct choice of basis, $\vectorstyle{\psi}$. Assume that $\vectorstyle{x}$ is composed of $n$ independently distributed random variables, $X_k$, with known supports and PDF, $f_{X_k}:\mathcal{X}_k\subseteq\mathbb{R}\rightarrow\mathbb{R}_{\geq0}$. The joint support, $\mathcal{X}$, and PDF, $f_{\vectorstyle{X}}$, are given by $\bigotimes_{k}\mathcal{X}_k$ and $\prod_{k}f_{X_k}$, respectively. Now recall that our goal is to propagate the input uncertainty on $\vectorstyle{x}$ to output $y$. By choosing the univariate bases, $\vectorstyle{\phi}^{(k)}$, so that they satisfy an orthogonality condition w.r.t. the PDFs, $f_{X_k}$, associated to the respective variables, $X_k$, the statistical moments can be calculated in function of the polynomial coefficients. 

Orthogonality of a basis is established in function of an inner product definition. We take interest in the inner product defined in (\ref{eq:improduct}). A polynomial basis is orthogonal w.r.t. $f_{X}$ if it satisfies the orthogonality condition $\langle \psi_i,\psi_j\rangle = \delta_{ij}\langle \psi_i^2\rangle$.
\begin{equation}
\label{eq:improduct}
\left\langle \psi_i,\psi_j\right\rangle \equiv \expect{\psi_i\psi_j} = \int_{\mathcal{X}} \psi_i(\vectorstyle{x}) \psi_j(\vectorstyle{x}) f_{\vectorstyle{X}}(\vectorstyle{x}) \text{d}\vectorstyle{x}
\end{equation}
Note that if we construct the multivariate basis as described above and so that the generating univariate bases, $\vectorstyle{\phi}^{(k)}$, satisfy the orthogonality condition w.r.t. to the PDFs, $f_{X_k}$, also the multivariate basis, $\vectorstyle{\psi}$, will satisfy the orthogonality condition w.r.t. the joint PDF, $f_{\vectorstyle{X}}$, considering that
\begin{equation*}
\begin{aligned}
\left\langle \psi_i,\psi_j\right\rangle &= \prod_{k=1}^{n}\int_{\mathcal{X}_k} \phi^{(k)}_{i_k}(x_k) \phi^{(k)}_{j_k}(x_k) f_{X_k}(x_k) \text{d}x_k \\
&= \prod_{k=1}^{n}\left\langle \phi_{i_k}^{(k)},\phi_{j_k}^{(k)}\right\rangle = \delta_{ij}
\end{aligned}
\end{equation*}

\subsubsection{Stochastic relation} Now recall that we desire to characterize the stochastic properties of the output by quantifying its stochastic moments, $\mu_m$
\begin{equation}
\label{eq:momentdef}
\mu_m = \expect{Y^m} = \expect{y(\vectorstyle{X})^m} = \int_\mathcal{X} y(\vectorstyle{x})^m f_{\vectorstyle{X}}(\vectorstyle{x}) \text{d}\vectorstyle{x}
\end{equation}

When we substitute the $d$-th order polynomial approximation of the output model in (\ref{eq:momentdef}), we retrieve an approximate expression for the $m$-th moment in function of the polynomial expansion coefficients.
\begin{equation}
\label{eq:moments}
\begin{aligned}
\mu^{(d)}_m &= \expect{y^{(d)}(\vectorstyle{X})^m} = \int_{\mathcal{X}} \left(\sum_{i\in\mathcal{I}} c_i \psi_i(\vectorstyle{x})\right)^m f_{\vectorstyle{X}}(\vectorstyle{x}) \text{d}\vectorstyle{x} \\
&= \sum_{i_1\in \mathcal{I}} \cdots \sum_{i_m\in \mathcal{I}} c_{i_1} \cdots c_{i_m} \left\langle\psi_{i_1}\cdots \psi_{i_m}\right\rangle
\end{aligned}
\end{equation}

From this expression one may easily verify that the first two moments can be estimated as shown below. Efficient estimation of the high-order moments is discussed in \ref{sec:algo}.
\begin{equation*}
\begin{aligned}
\mu_1^{(d)} &= c_1 \langle\psi_1^2\rangle\\
\mu_2^{(d)} &= \smallsum_{i\in\mathcal{I}} c^2_i \langle\psi_i^2\rangle
\end{aligned}
\end{equation*}

We emphasize that this result only holds when the series is expanded over a basis that satisfies the orthogonality condition w.r.t. $f_{\vectorstyle{X}}$. Several standard probability distributions are associated to known polynomial families by the Wiener-Askey scheme \cite{xiu2002wiener}. An overview is presented in Table \ref{tab:wiener}. If the input stochasticity does not correspond with a standard distribution, a variable transformation can be used.

\subsubsection{Variable transformation}
A set of models with parameterized uncertainty parameters $\vectorstyle{X}$ can now be characterized. To that end we introduce the parametric transformation, $\vectorstyle{x}(\cdot|\vectorstyle{\alpha})$, that maps standard random variable, $\vectorstyle{\Theta}\in\vartheta$, with entries that are distributed according to one of the distributions in Table \ref{tab:wiener}, to the random variable, $\vectorstyle{X}$. The multivariate expansion basis, $\vectorstyle{\psi}$, can then be generated from the corresponding standard polynomials. We emphasize that the forward model becomes a function of the variable, $\vectorstyle{\Theta}$. 
Consequently, the conditional PDF of the outcome is now fully determined by parameter $\vectorstyle{\alpha}$, and the distribution of $\vectorstyle{\Theta}$.
\begin{equation}
\label{eq:mapping}
\vectorstyle{X} = \vectorstyle{x}(\vectorstyle{\Theta}|\vectorstyle{\alpha}) \rightarrow Y = y\left(\vectorstyle{x}\left(\vectorstyle{\Theta}|\vectorstyle{\alpha}\right)\right) \equiv \eta\left(\vectorstyle{\Theta}|\vectorstyle{\alpha}\right)
\end{equation}

\begin{table}[t]
	\caption{\label{tab:wiener}Wiener-Askey polynomial chaos.}
	\centering
	\begin{tabular}{c c c}
		\toprule
		distribution, $f_{X}$ & polynomials, $\phi_i$ & support, $\mathcal{X}$ \\
		\midrule
		\midrule
		Gaussian & Hermite & $\left[-\infty,\infty\right]$ \\
		Gamma & Laguerre & $\left[0,\infty\right]$ \\
		Beta & Jacobi & $\left[-1,1\right]$ \\
		Uniform & Legendre &  $\left[-1,1\right]$ \\
		\bottomrule
	\end{tabular}
	\vspace*{-10pt}
\end{table}

\subsubsection{Coefficient determination} The polynomial coefficients from (\ref{eq:poly-D}) can be quantified numerically \cite{xiu2007efficient} by projection of the forward model on the polynomial space exploiting the properties of the inner product. All terms but one will vanish resulting in 
\begin{equation}
\label{eq:galerkin}
c_{i}(l|\vectorstyle{\alpha}) = \left\langle \eta(l,\vectorstyle{\Theta}|\vectorstyle{\alpha}),\psi_i\right\rangle = \int_{\vartheta} \eta(l,\vectorstyle{\theta}|\vectorstyle{\alpha})\psi_i(\vectorstyle{\theta})f_{\Theta}(\vectorstyle{\theta})\text{d}\vectorstyle{\theta}
\end{equation}

for all $i\in \mathcal{I}$. Since the scope of this work is on general nonlinear forward models, we can not evaluate the associated integral explicitly. We therefore approximate the integrals using Gaussian-quadrature (\ref{eq:quadrature}). For details we refer to \cite{xiu2005high}. 

An univariate Gaussian-quadrature of order $q$ is defined as a signature of the same size. A signature is defined as a set $\{(w_j,\vectorstyle{\theta}_j)\}_{j\in\mathcal{Q}},\mathcal{Q}=\{1,\dots,q\}$, where $\vectorstyle{\theta}_j$ and $w_j$ are associated to the position and the weight attributed to each element, respectively. The signature will depend on the weighting function, $f_{\Theta}$, and is as such related to the polynomial basis. A quadrature of order $q$ is exact for polynomials up to degree $ 2q-1$. The order hence affects the number of coefficients that can be retrieved correctly and therefore the accuracy of the polynomial approximation. In the multivariate case, a full tensor product of univariate quadrature rules can be considered. We note that the number of collocation points will thus scale exponentially with the number of dimensions. 
\begin{equation}
\label{eq:quadrature}
c_{i}(l|\vectorstyle{\alpha}) \approx \sum_{j\in\mathcal{Q}} \eta\left(l,\vectorstyle{\theta}_j|\vectorstyle{\alpha}\right)\psi_i\left(\vectorstyle{\theta}_j\right)w_j
\end{equation}

Note that, similar to MC techniques, the gPC approach requires a number of evaluations of the forward model, $\eta(l,\vectorstyle{\theta}_j|\vectorstyle{\alpha})$. The difference is that the estimation of the statistical moments is realized through a mathematical detour. The focus of approximation in the gPC framework is on modeling the polynomial response function whilst that of the MC approach is on the direct estimation of the output distribution. The accuracy of gPC depends on the capacity of the basis to capture the nonlinearity of the forward model rather than on the capacity of the sampling method to properly represent the input uncertainty by the spatial distribution of the sample points. It is the prevailing consensus that an equivalent level of accuracy can be achieved with only a fraction of the input points of any MC approach.

\subsection{PDF fitting with the method of moments}
\label{sec:PDF-MM}
Once the approximate stochastic moments, $\mu_{l,m}^{(d)}(\vectorstyle{\alpha})$, of the $l$-the experiment for given input stochastic model, $\vectorstyle{\alpha}$, are obtained a model for the conditional PDF of the random outcome, $Y_l$, can be extracted by matching these moments to a known parametric PDF, the so called methods of moments. The complexity of the parametric PDF should match the stochasticity of the random input, the expected symmetry of the stochasticity of the outcome and the nonlinearity of the forward model. Two possibilities are discussed for extracting the conditional PDF.
\begin{itemize}
	\item{\textit{Gaussian distribution}:} The most straightforward and prevailing approach is to fit a normal distribution to the output PDF based on estimates of the mean and variance. The conditional PDF is then approximated by
	\begin{equation*}
	f_{Y}(y|\vectorstyle{\alpha}) = \varphi\left(\frac{y - \mu^{(d)}(\vectorstyle{\alpha})}{\sigma^{(d)}(\vectorstyle{\alpha})}\right)
	\end{equation*}
	
	\item{\textit{Maximum entropy distribution}:} To cope with an increased number of moments, one can fit a maximum entropy distribution. Here, we seek a distribution $f_{Y}$ that maximizes the entropy $\operatorstyle{W} =- \int_{\mathcal{Y}} f_Y(y) \log f_Y (y) \text{d}y$ subject to $M+1$ moment matching constraints. The solution to this optimization problem is given by the following parameterized PDF (see Appendix \ref{appendix:med} for derivation)
	\begin{equation*}
	f_{Y}(y|\vectorstyle{\lambda}) = \exp \left(-\mathsmaller{\sum}_{m=0}^M\lambda_{m} y^m\right)
	\end{equation*}
	
	Substituting the latter into the original constrained optimization problem, yields the following dual unconstrained optimization problem that can be solved to find values for the parameters $\lambda_m$
	\begin{equation*}
	\max_{\vectorstyle{\lambda}}  \log \int_{\mathcal{X}} \exp \left(-\mathsmaller{\sum}_{m=1}^M \lambda_m x^m\right) \text{d}x + \mathsmaller{\sum}_{k=1}^M \lambda_m \mu^{(d)}_{m}(\alpha)
	\end{equation*}
	A solution exists for any $M$. However, in order to capture fourth order asymmetric effects (i.e. bimodal densities), the number of moments should $\geq 4$ so that the polynomial in the exponent can have two local maxima.	
\end{itemize}

Worth mentioning is the common conception in the literature \cite{lee2009comparative,rajabi2015polynomial} to obtain the moments, or even directly the PDF, by applying MC techniques on the polynomial expansion (\ref{eq:poly-D}) which is relatively cheap to evaluate. In this approach the expansion is used as a response function and the benefit of gPC is only exploited through the supposed superior convergence rate of the expansion for polynomials corresponding the input distribution (\ref{eq:poly-D}). 
When compared to its peers, gPC is usually applied in this sense \cite{xiu2005high,lee2009comparative,najm2009uncertainty}.

\vspace*{-10pt}
\subsection{Estimation of high-order moments}
\label{sec:algo}
The expression presented in (\ref{eq:moments}) becomes inefficient for $m>2$. Therefore we expand the power of the sum by applying the multinomial theorem to obtain an expression which is numerically far more efficient. Remark that there is only one summation and each term in the series is rendered unique. Moreover, we decompose the high-order multivariate inner products into a product of univariate inner products. As a result we can compute the high-order inner products numerically using a univariate instead of a multivariate quadrature. The multi-index set $\mathcal{I}_{m,p}$ is defined as $\{\uline{i}\in\mathcal{I}_{m,p}:|\uline{i}|=m\}$.
\begin{equation}
\label{eq:moments2.0}
\begin{aligned}
\mu_m^{(d)} = \sum_{\uline{i} \in\mathcal{I}_{mp} } \binom{m}{i_1 \cdots i_p} \cdot \prod_{j=1}^n \left\langle\prod_{k=1}^p {\phi_{\uline{k}(j)}^{(j)}}^{i_k}\right\rangle \cdot \prod_{k=1}^p c_k^{i_k} 
\end{aligned}
\end{equation}
Details on deriving equation (\ref{eq:moments2.0}) can be found in Appendix \ref{appendix:mathderiv}. The algorithmic execution time still demands an upper-limit for $m$. That is because $\dim(\mathcal{I}_{m,p})=\binom{m+p-1}{m}$ where $p$ itself is a combinatorial number. Note that the orthogonality property extends to the high-order inner products, as they occur in (\ref{eq:moments}) or (\ref{eq:moments2.0}) for $m>2$, however in a less predictable fashion. 

\begin{figure}[t!]
	\centering
	\begin{subfigure}[h]{.20\columnwidth}
		\caption{$\mathcal{I}_{5,10}$}
		\includegraphics[trim=8.5cm 10cm 8.5cm 9.5cm,clip=true,width=\columnwidth]{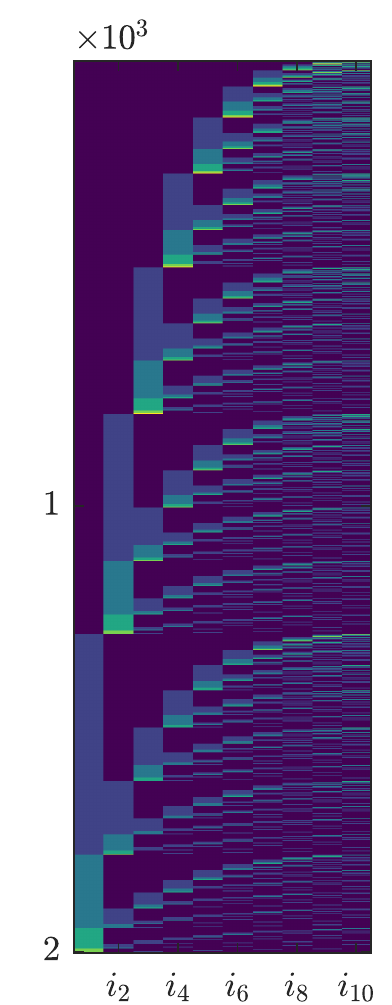}
	\end{subfigure}
	\quad
	\begin{subfigure}[h]{.2368\columnwidth}
		\caption{$\mathcal{I}^*_{5,10}$}
		\includegraphics[trim=7.75cm 10cm 8.5cm 9.5cm,clip=true,width=\columnwidth]{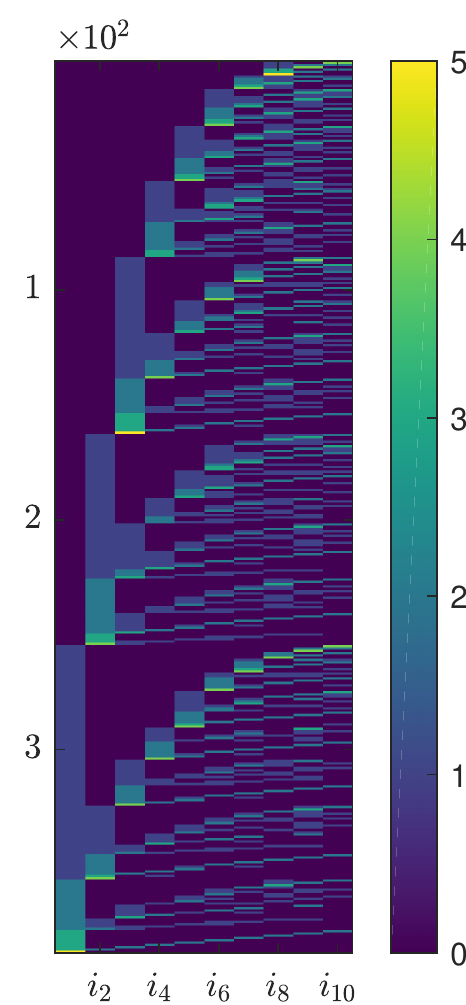}
	\end{subfigure}
	\quad
	\begin{subfigure}[h]{.375\columnwidth}
		\includegraphics[width=\columnwidth]{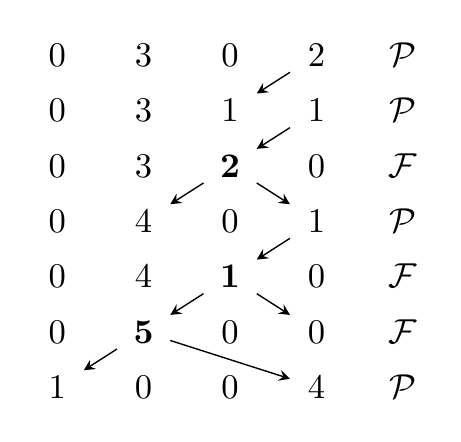}
	\end{subfigure}
	\caption{\textit{Left}: Illustration of the sets $\mathcal{I}_{5,10}$ (\textit{left}) and $\mathcal{I}^*_{5,10}$ (\textit{right}) for $(n,d)=(2,3)\rightarrow p = 10$. \textit{Right}: Illustration of the calculation mechanism of $\mathcal{I}_{m,p}$ for $(m,p)=(5,4)$.}
	\label{fig:sets}
	\vspace{-10pt}
\end{figure}

Hence in practice we only require the multinomial powers $\uline{i}$ whose corresponding inner product is nonzero. Such are stored in the set $\mathcal{I}^*_{m,p}$. The corresponding multinomial coefficient and inner product are compressed into coefficients $a_{\uline{i}}^{n,d,m}$ and stored for later use. One may verify that a significant fraction of $\mathcal{I}_{m,p}$ result into zero valued inner products, see Fig. \ref{fig:sets}.
\begin{equation}
\label{eq:moments3.0}
\begin{aligned}
\mu_m^{(d)} = \sum_{\uline{i} \in \mathcal{I}^*_{m,p}} a_{\uline{i}}^{n,d,m} \prod_{k=1}^p c_k^{i_k} 
\end{aligned}
\end{equation}

To compute the set $\mathcal{I}_{m,p}^*$, we devised an algorithm that calculates the next multi-index, satisfying $|\uline{i}|=m$, from the previous multi-index. Because our algorithm generates the admissible multi-indices one by one, it admits further increase of the computational efficiency. The process of generating the multi-indices and the corresponding inner product can be executed in parallel avoiding storage of the entire set $\mathcal{I}_{m,p}$. 

\begin{figure}[b!]
	\vspace{-10pt}
	\centering
	\includegraphics[width=.5\columnwidth]{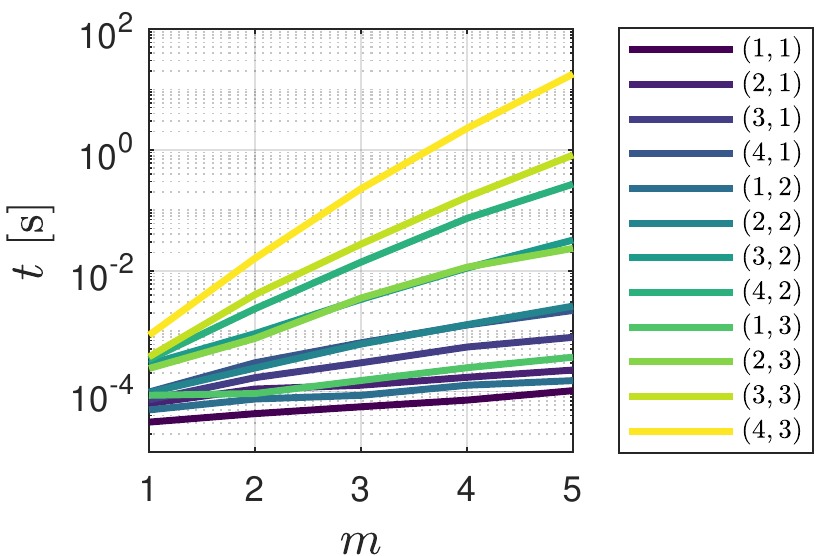}
	\caption{Calculation time of the offline step w.r.t. $(n,d,m)$.}
	\label{fig:zeroprod}
\end{figure}

The general idea is presented in Fig. \ref{fig:sets}. The multinomial power problem can be addressed as reorganizing a stack of $m$ items over $p$ possible locations. We use two logical operators, push ($\mathcal{P}$) and fork ($\mathcal{F}$). A row is `pushed' if the last location is nonzero. Otherwise the trailing nonempty location is `forked' which involves pushing $1$ item from this location to the next and placing the remaining items back to the last location. The algorithm is initiated with all items in the last position.  The calculation time of the parallel offline step is shown in Fig. \ref{fig:zeroprod} and could be limited, e.g. $(n,d,m)=(4,3,5)$, to $18\si{\second}$  nonetheless that $575,757$ inner products are validated. The generation of the multi-indices itself took only about $1.2 \si{\second}$. Calculations were performed on a 2.1GHz, 4 GB RAM laptop.

Above procedure allowed us to compute up to the 5\textsuperscript{th} moment quite efficiently. As discussed in section \ref{sec:PDF-MM}, this allows us to fit bimodal output distributions rigorously.

\subsection{Forward uncertainty propagation}
\begin{figure}[t!]
	\centering
	\includegraphics[trim=5.25cm 9.5cm 4cm 10cm,clip=true,width=.65\columnwidth]{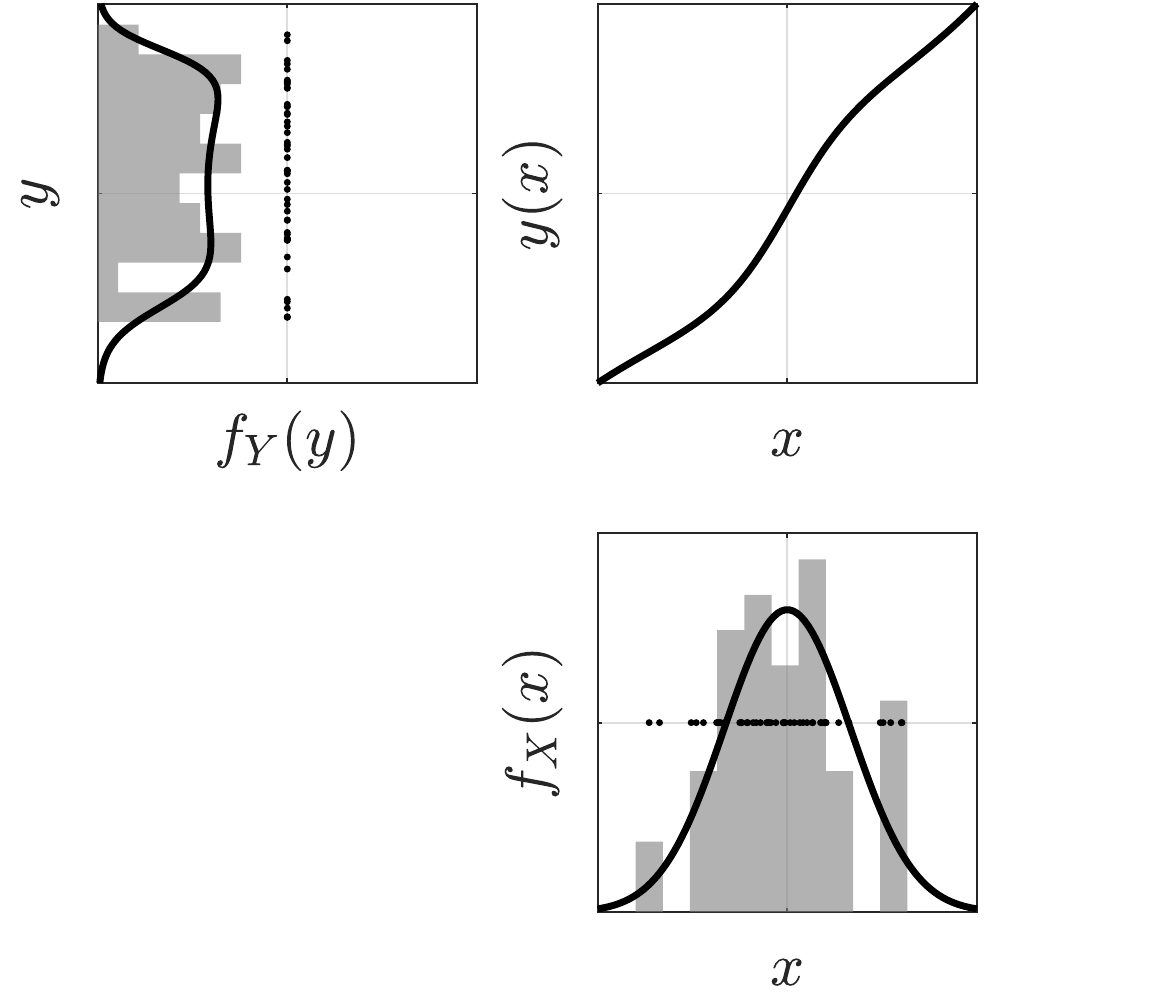}
	\caption{Nonlinear transformation of standard normal random variable. The lower right plot shows the density of the original random variable, $X$. The variable is passed through the function displayed in the upper right. The density of the output random variable, $Y$, is plotted in the upper left graph.}
	\label{fig:PDFtransform}
	\vspace{-3pt}
\end{figure}

In order to illustrate the forward propagation of uncertainty concatenating the techniques described in the sections \ref{sec:UP-PCE} to \ref{sec:algo}, we apply the methodology on an illustrative univariate nonlinear forward model, $y(x)$. 
\begin{align*}
y(x) &= \tan\left(\tfrac{1}{4}x\right) + \exp\left(\tfrac{1}{3}x-1\right) + \tanh(x)
\end{align*}
For convenience we further assume that the random variable, $X$, is distributed according to the standard normal distribution, $\mathcal{N}(0,1)$. The forward propagation of this random variable passed through the function $y(x)$, results into a \textit{camelback} output distribution as is depicted in Fig. \ref{fig:PDFtransform}. The lower right plot shows the density of the stochastic input, $X$. The density of the output random variable, $Y$, is plotted in the upper left graph. Additionally, an MC sample set of $N=50$ is visualized by the black dots with corresponding $10$ bin histogram in gray.

\begin{figure}[t!]
	\centering
	\begin{subfigure}[b]{.9\columnwidth}
		\caption{Polynomial coefficients,  $c_i [\cdot]$}
		\vspace*{-3pt}
		\includegraphics[trim=3.5cm 12.25cm 4cm 13cm,clip=true,width=\columnwidth]{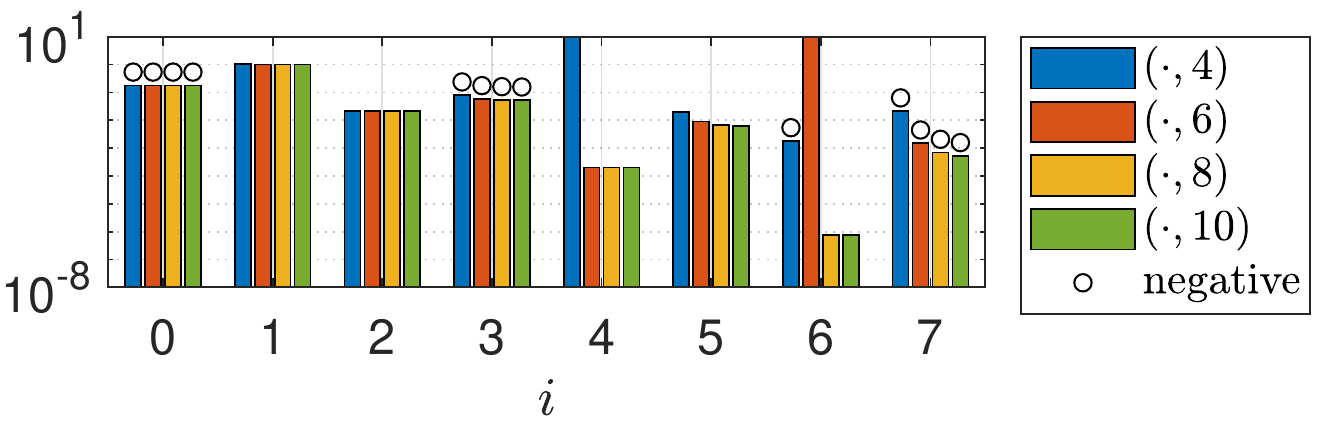}
		\label{fig:coeff}
	\end{subfigure}
	\begin{subfigure}[b]{.9\columnwidth}
		\vspace*{-10pt}
		\caption{Stochastic moments discrepancy, $\Delta \mu_i [\%]$}
		\vspace*{-3pt}
		\includegraphics[trim=3.5cm 12.25cm 4cm 13cm,clip=true,width=\columnwidth]{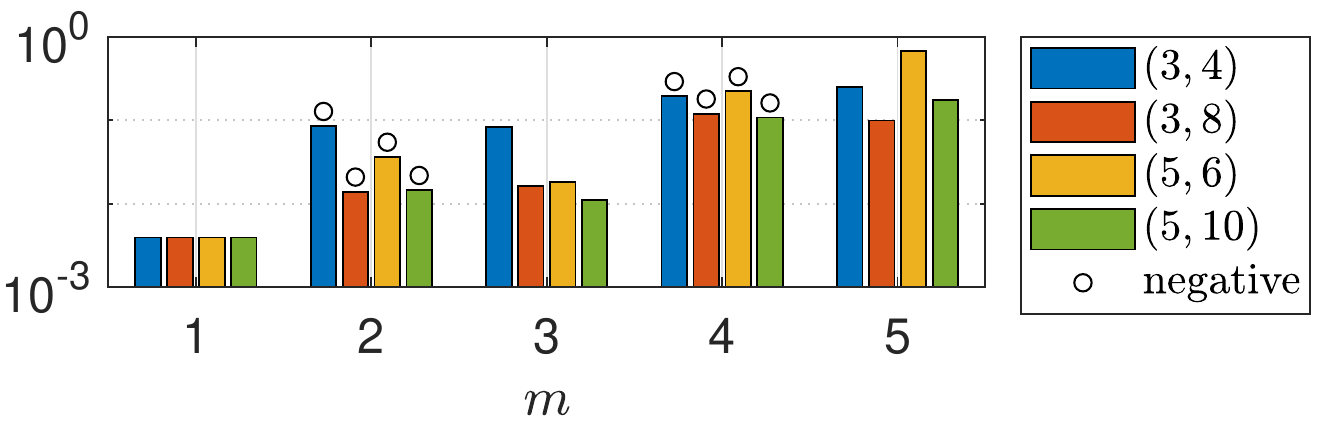}
		\label{fig:moments}
	\end{subfigure}
	\vspace*{-15pt}
	\caption{Polynomial chaos coefficient estimates for varying quadrature orders are displayed in the top graph. Stochastic moments discrepancy w.r.t. the true moments as obtained from the gPC coefficients and by MC sampling of the gPC approximations are displayed in the bottom graph. Negative values are indicated by a white marker.}
	\label{fig:coeff-moments}
	\vspace{-3pt}
\end{figure}

Figure \ref{fig:coeff} depicts the polynomial coefficients for varying quadrature order, $q$, and using Hermite chaos. One may observe that the higher order coefficients are poorly approximated with low order quadratures. Recall that a quadrature of order $q$ is exact for polynomials up to degree $ 2q-1$. Now, assuming that model $y(x)$ contains polynomials up to degree $d$ and that we desire to approximate the $d$-th coefficient, then the integrand in (\ref{eq:galerkin}) will contain a polynomial of degree $2d$. In order to be exact, the quadrature should therefore have order $d+1$ at least. Remark that the coefficient estimate becomes increasingly precise for $q$ surpassing $d+1$, as can be seen for e.g. $i=3$. That is because, $y(x)$ is truly a polynomial of degree $\gg d$. 

In Fig. \ref{fig:moments} we compare the stochastic moments numerically obtained from the true PDF\footnote{For a monotonic function, $y(x)$, with inverse, $x(y)$, this is equal to, $f_Y(y) = \lvert x'(y) \rvert \cdot f_X(x(y)) = \lvert y'(x(y)) \rvert^{-1} \cdot f_X(x(y)) $.} with those obtained from gPC approximation (\ref{eq:poly-D}) for varying $d$ and $q$. Note that we have chosen the quadrature order so that $q\geq d+1$. As a result, the poorly approximated coefficients from Fig. \ref{fig:coeff} are never considered in the moment computation. This explains why even for low polynomial order, the higher order moment approximations remain very precise.

\begin{figure}[t!]
	\centering
	\begin{subfigure}[b]{.47\columnwidth}
		\caption{Gaussian}
		\vspace*{-5pt}
		\includegraphics[trim=3.9cm 9.7cm 5cm 10.8cm,clip=true,width=\columnwidth]{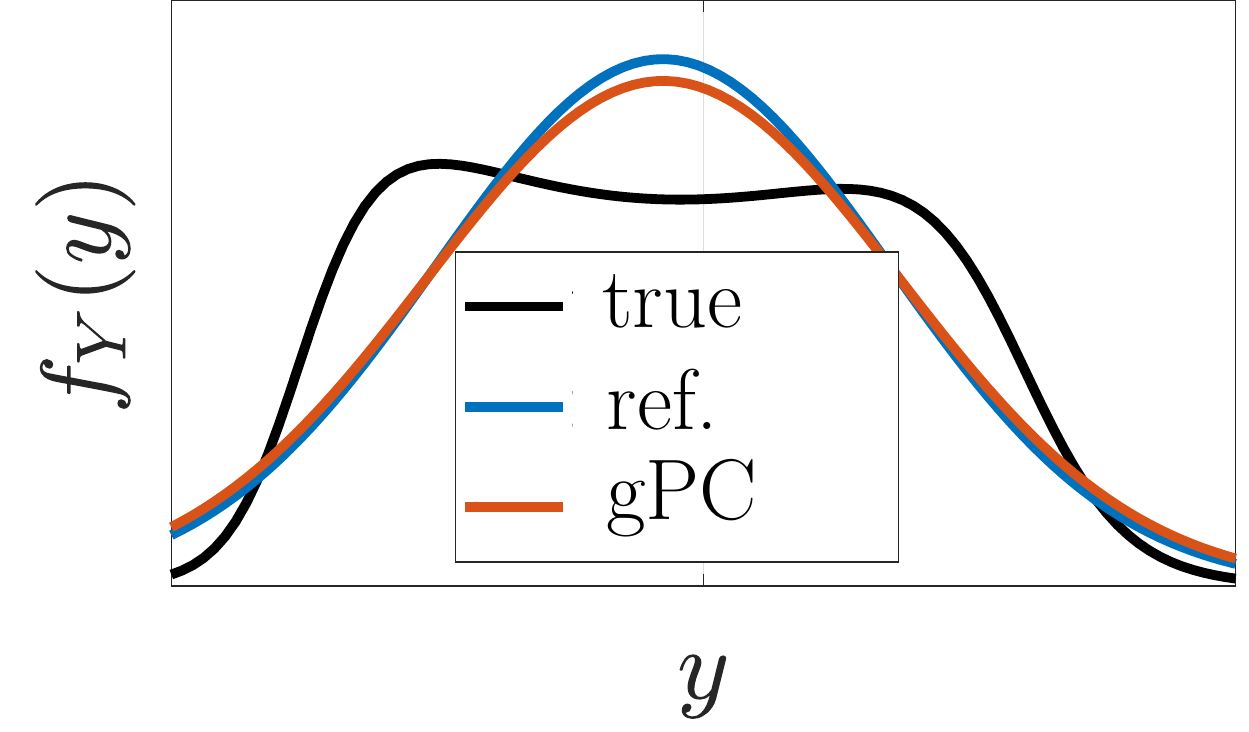}
	\end{subfigure}
	\begin{subfigure}[b]{.47\columnwidth}
		\caption{Maximum entropy}
		\vspace*{-5pt}
		\includegraphics[trim=3.9cm 9.7cm 5cm 10.8cm,clip=true,width=\columnwidth]{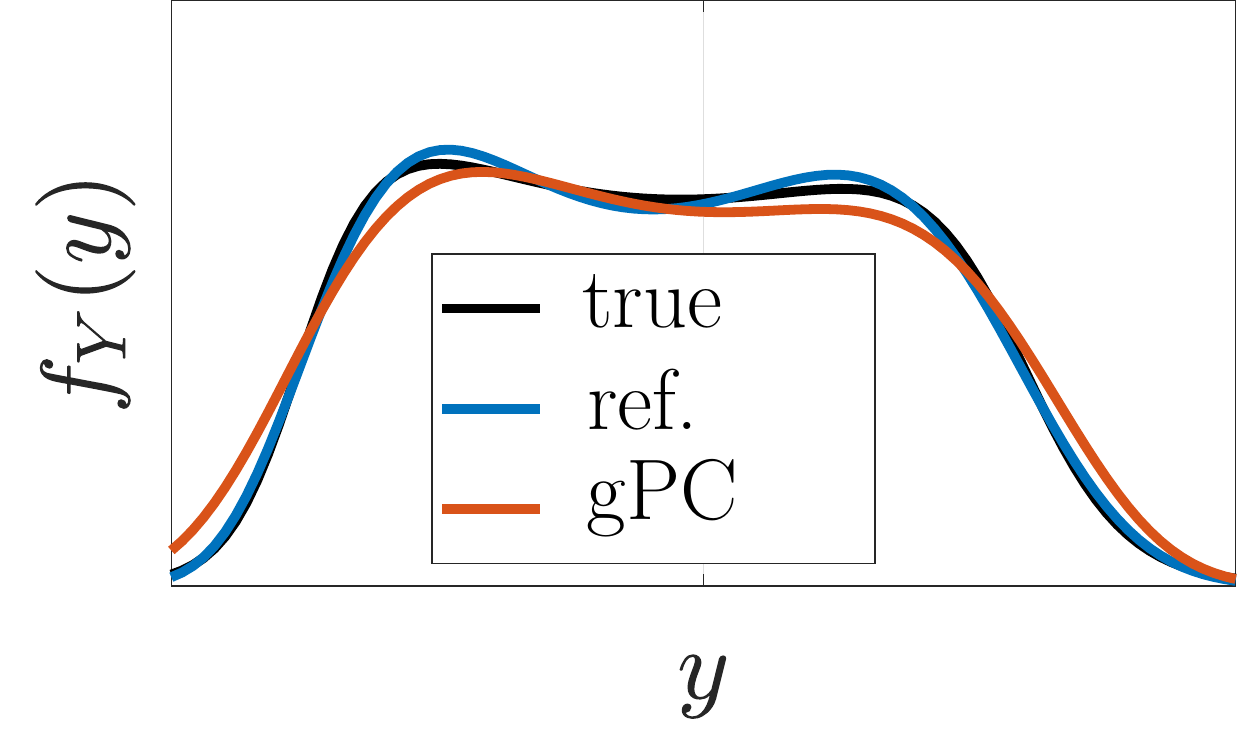}
	\end{subfigure}
	
	\begin{subfigure}[b]{.475\columnwidth}
		\includegraphics[trim=5cm 13cm 5.75cm 13cm,clip=true,width=\columnwidth]{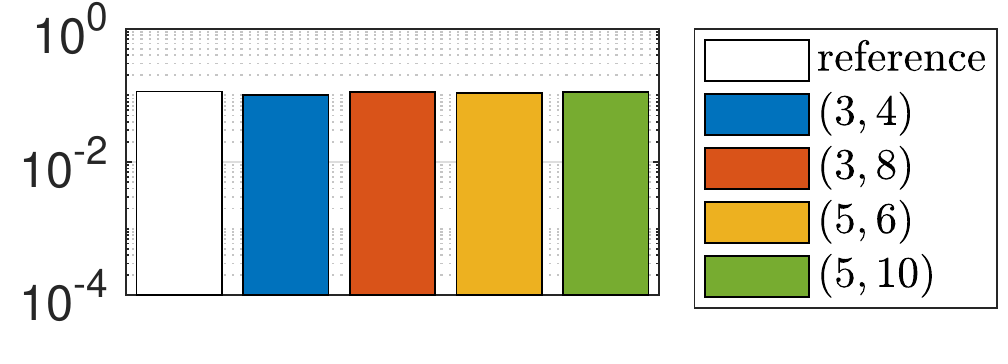}
	\end{subfigure}
	\begin{subfigure}[b]{.475\columnwidth}
		\includegraphics[trim=5cm 13cm 5.75cm 13cm,clip=true,width=\columnwidth]{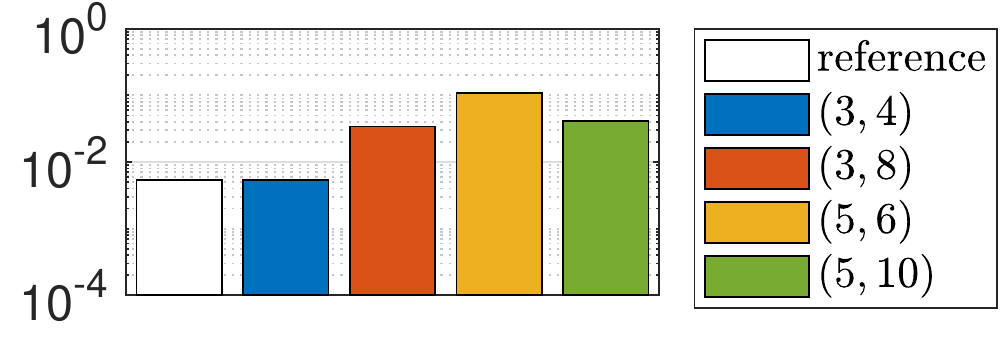}
	\end{subfigure}
	\caption{Comparison of PDF approximations of a Gaussian, a two component mixture and a maximum entropy distribution model fit. The top three figures portray the PDFs, corresponding EMD values are compared in the bottom three.}
	\label{fig:matching}
\end{figure}

In conclusion, the moments are employed to fit parametric distributions to the output distribution, $f_Y$. Results are depicted in Fig. \ref{fig:matching} when using the Gaussian and maximum entropy distribution. The numerical correspondence between two PDFs is quantified by the Earth Mover's Distance (EMD) (Appendix \ref{appendix:emd}). Each fit is compared to the actual PDF. The PDF fit obtained with the exact moments serves as an upper bound for the amount of information that is contained within the moments.

\section{Wet clutch shifting time}
\label{sec:application}

\subsection{System description}
\begin{figure}[h!]
	\centering
	\includegraphics[width=0.6\columnwidth]{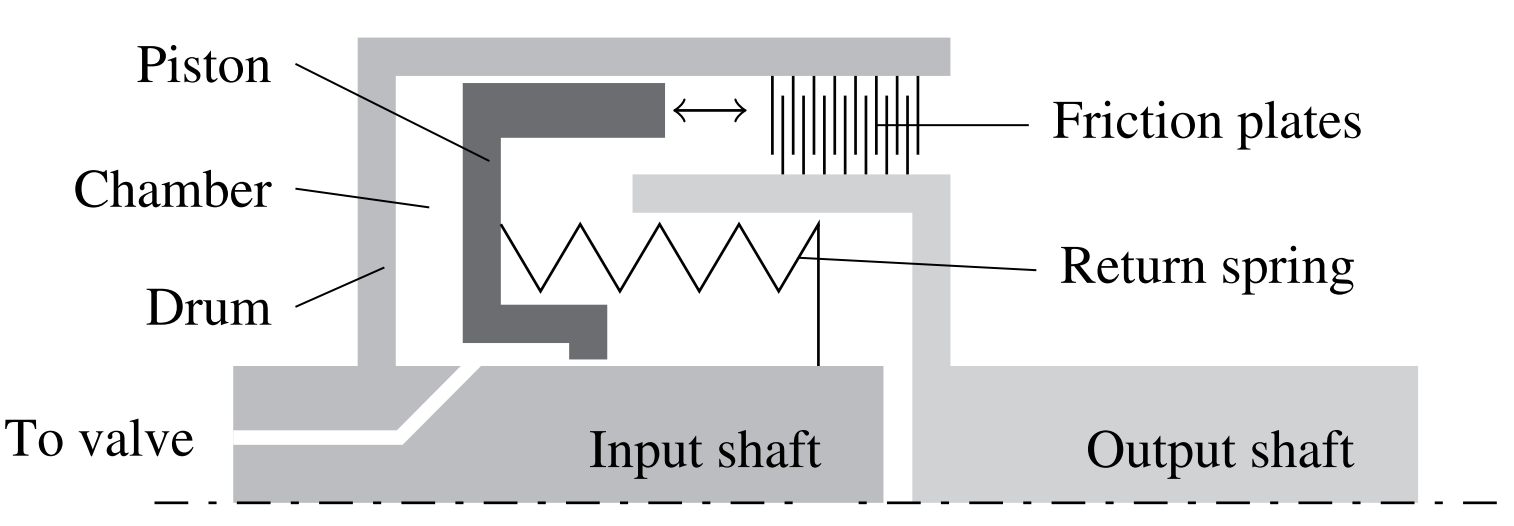}
	\caption{Schematic cross section of the wet clutch system.}
	\label{clutch}
\end{figure}

\begin{figure}[h!]
	\vspace{-4pt}
	\centering
	\includegraphics[width=0.45\columnwidth]{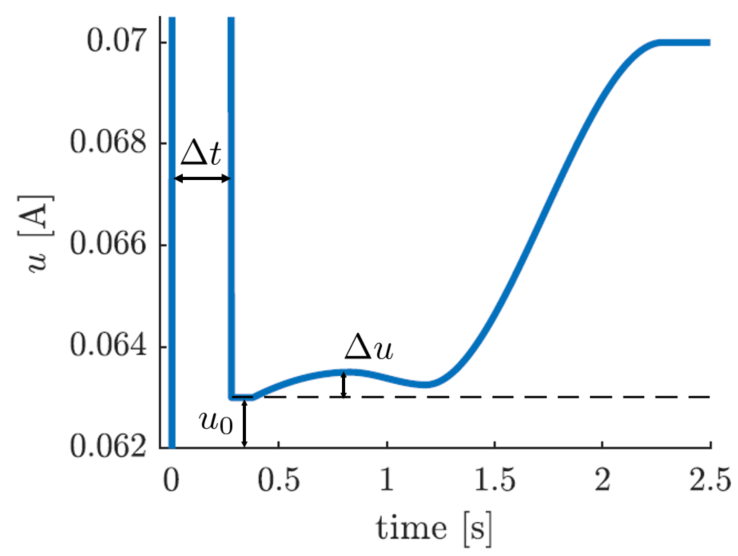}
	\caption{Feedforward current signal.}
	\label{fig:feedforward}
	\vspace{-10pt}
\end{figure}

The proposed methodology is applied to identify the parametric uncertainty of a wet clutch system. A clutch is a mechanical device that connects a motoring unit to a load that transforms power into useful work. The objective is to ensure smooth and on demand coupling of the driving shaft to the driven shaft. A wet clutch system realizes this by means of contact friction between two series of friction plates, see Fig. \ref{clutch} for a schematic cross section \cite{bruno2011}. By construction these friction plates are fully submerged in oil so that slip induced heat can be transferred adequately. This avoids thermal overheat and increases the clutch's life time. A wet clutch uses the same hydraulic system to activate the engagement process. By pressurizing the hydraulic chamber, a freewheeling piston is pressed against the series of interlocking friction plates realizing a power transfer. The pressure in the hydraulic chamber is controlled via a current signal that activates a servo-valve linking the hydraulic chamber to a pressurized reservoir. The pressure difference over the piston determines its position and therewith the clutch engagement. 
Since the wet-clutch set-up should be low cost, additional sensors are avoided and a feedforward control strategy using the parameterized current profile shown in Fig. \ref{fig:feedforward} is applied. The signal is characterized by three parameters. An initial maximum current pulse, parameterized by its duration $\Delta t$, gets the piston as close as possible to the friction plates without making contact.  During this phase, the hydraulic chamber fills up and initial torque transfer commences due to Couette flow phenomena. The pressure drop invoked by reducing the current to $u_0$ avoids a hard collision. Once the plates are in contact a second pulse, defined by $\Delta u$, is generated to have a proper pressure build-up for synchronizing the load. At synchronous speed the torque transfer caused by slip speeds drops away. Therefore, hazardous jerks should be avoided by having slow synchronization. This explains the downward flank of the current pulse. After syncing the pressure is increased to guarantee the plates cannot release again by accident and start slipping. The shifting time, defined as the time interval between initialization of the feedforward control signal and the moment that both shafts obtain the same angular speed, is therefore considered as key characteristic of the engagement process.

\subsection{Shifting time model}
\label{sec:model&exp}
\begin{figure}[h!]
	\centering
	\includegraphics[width=0.8\columnwidth]{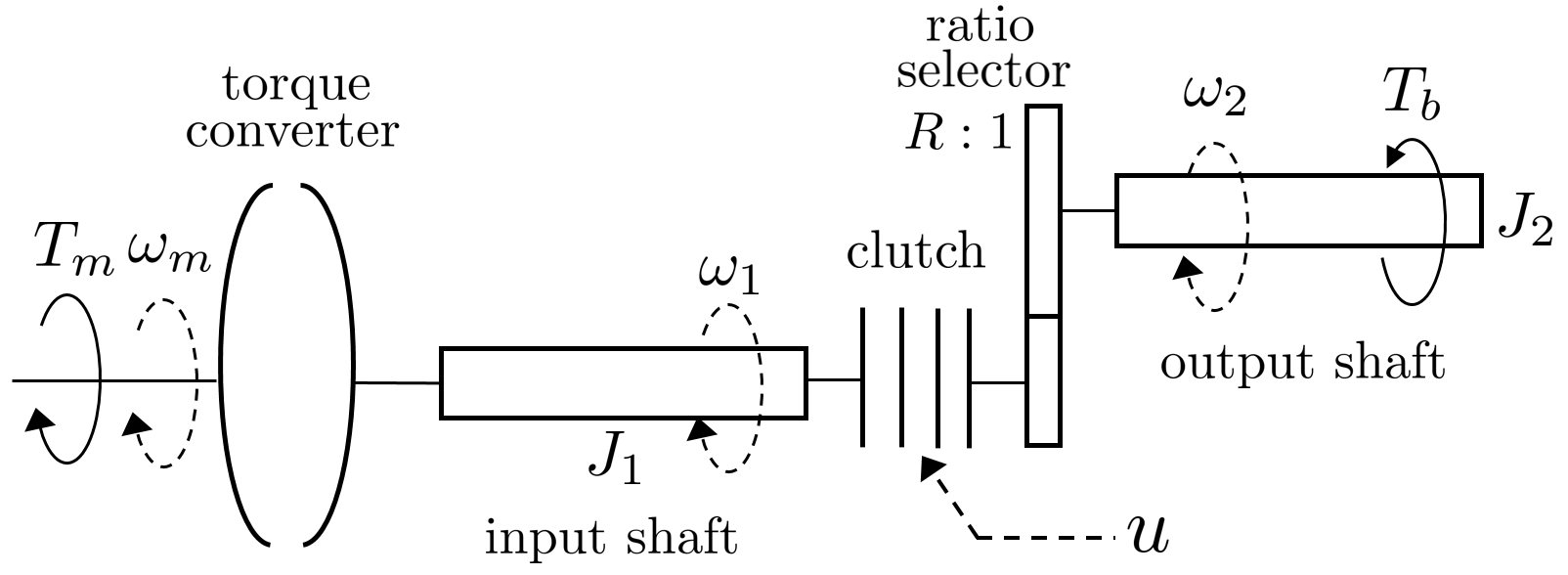}
	\caption{Dynamic torque equilibrium scheme.}
	\label{DynamicScheme}
\end{figure}

The shifting time is derived from a dynamic simulation of the clutch engagement. Prior research was on the building of a dynamic model of the engagement process based on first principle physics. Model parameter values were then identified via experimental system identification (from isolated subsystem testing....). The exact model identification is not in the scope of this paper and a general overview is given. We refer to \cite{Georges,iqbal2015,agusmian2013,Widanage2011} for further reading. Figure \ref{DynamicScheme} illustrates the simplified dynamical scheme of the wet clutch system.  
 The motor is controlled towards a constant speed $\omega_m$ delivering a torque $T_m$ to the system. The motor is connected with a torque converter that drives the input shaft of the clutch with torque $T_1$ at rotational velocity $\omega_1$. The system can be modeled using a nonlinear torque ratio function $\frac{T_1}{T_m} = f_{t}(\frac{\omega_1}{\omega_m})$ and a capacity factor function $\frac{\omega_{m}}{\sqrt{T_m}}=f_{c}(\frac{\omega_1}{\omega_{m}})$ provided by the manufacturer. Consequently, the driving torque $T_1$ of the input shaft can be written as 
\begin{equation*}
T_1(\omega_m,\omega_1) = \omega_m^2\frac{f_{t}(\frac{\omega_1}{\omega_m})}{f_{c}(\frac{\omega_1}{\omega_m})^2}. \\
\end{equation*}
The clutch is connected to a ratio selector with gear ratio $R=\frac{\omega_{in}}{\omega_{out}}$. Thereafter, the output shaft operates at $\omega_2$ as result of the driving torque $T_2$. Furthermore, the rotational movement of the output shaft and flywheel, characterized by inertia $J_2$, are counteracted by an additional brake torque $T_b$. 
\begin{equation*}
T_b(\omega_2)=T_{b0} + b\omega_o
\end{equation*}
The system is fully engaged when both shafts have an equivalent speed $\omega_1=R \omega_2$. The dynamics of this engagement process can be described by a succession of two distinct phases.\\

\subsubsection{Asynchronous phase}
\begin{figure}[h!]
	\centering
	\begin{subfigure}[b]{.45\columnwidth}
		\caption{pressure}
		\vspace*{-5pt}
		\includegraphics[trim=6.5cm 11.9cm 7.5cm  12cm,clip=true,width=1\columnwidth]{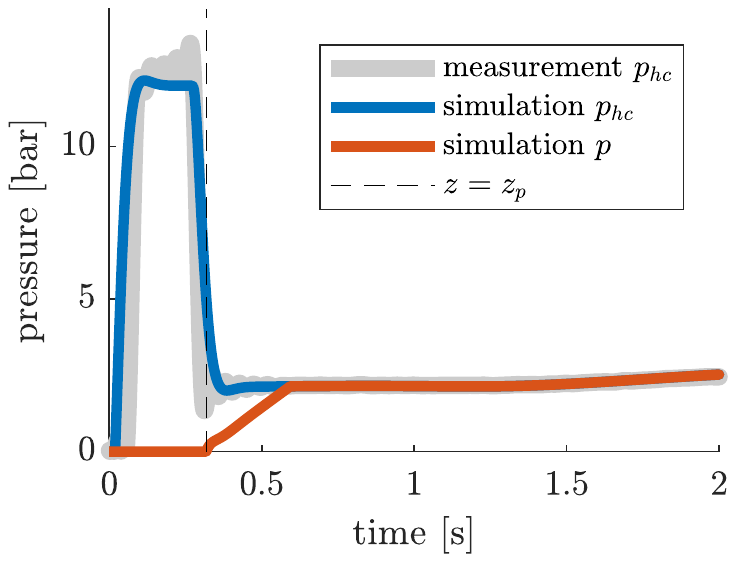}
		\label{fig:pressure}
	\end{subfigure}
	\begin{subfigure}[b]{.45\columnwidth}
		\caption{piston position}
		\vspace*{-5pt}
		\includegraphics[trim=6.5cm 11.9cm 7.5cm  12cm,clip=true,width=1\columnwidth]{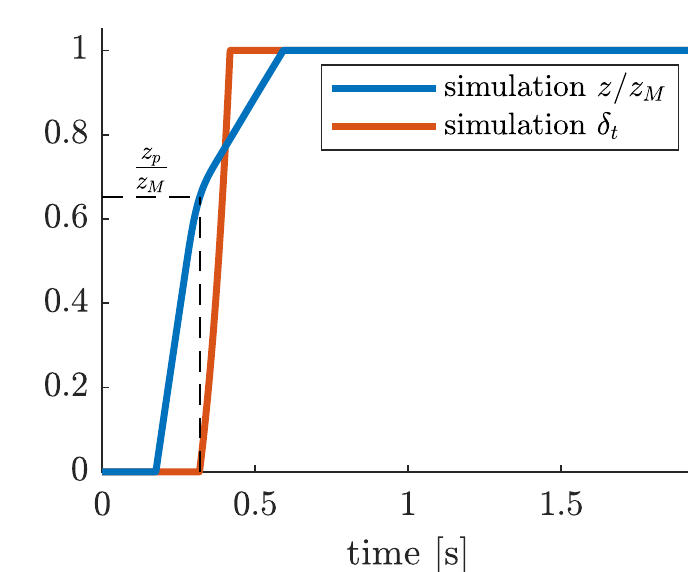}
		\label{fig:pos}
	\end{subfigure}
	\caption{Illustration of overpressure and resulting piston position.}
	\label{fig:pressAndPosition}
	\vspace{-5pt}
\end{figure}

The first phase is characterized by a speed difference $\omega_1 > R\omega_2 $ of the clutch shafts. During this phase the hydraulic piston chamber starts filling up. The resulting overpressure $p_{hc}$ is measured and depicted in Figure \ref{fig:pressure}. The dynamics of the pressure build-up can be captured by defining a state $\vectorstyle{s}=\begin{bmatrix}
p_{hc}&\dot{p}_{hc}&\tilde{z} \end{bmatrix}^T$. The state variable $\tilde{z}$ can easily be scaled and truncated within a feasible area to obtain the piston position $z\in[0, z_M]$, for which $z_M$ is the position that assures full contact between the friction plates. The dynamics are modeled by an affine system with parameters identified to optimally match the pressure $p_{hc}$. Mark that the control input $u$ is the feedforward current signal of the valve as was illustrated in Fig. \ref{fig:feedforward}.
\begin{equation}
\begin{aligned}
\label{eq: pressureDyn}
\dot{\vectorstyle{s}}&=\matrixstyle{A}  \vectorstyle{s}+ \matrixstyle{B}  \begin{bmatrix}
u \\ \dot{u}
\end{bmatrix}+\vectorstyle{c} \\
\matrixstyle{A}&=\begin{bmatrix}
0  & 1 & 0 \\ a_1 & a_2 & 0 \\ a_3 & a_4 & a_5\\
\end{bmatrix};
 \ \vectorstyle{B}=\begin{bmatrix}
0 & 0 \\ b_1 & b_2  \\ 0 & 0 \\ 
\end{bmatrix}; \ 
\vectorstyle{C}=\begin{bmatrix}
0  \\ c_1 \\ c_2 
\end{bmatrix}
\end{aligned}
 \end{equation}
The detailed pressure dynamics are out of scope and we refer to prior research for more information \cite{Widanage2011,Georges}. During the filling phase the plates do not make contact. However, the oil between the friction plates will behave as a planar Couette flow. The resulting shear stresses within the fluid will cause an initial torque transmission and acceleration of the load speed $\omega_2$. This phenomena is characterized by the constant $\gamma $, capturing the geometry of the plates and the fluid viscosity. A second influence on the torque transmission is the pressure on the plates $p$. This pressure is a fraction of the total hydraulic overpressure $p_{hc}$ as is illustrated in Fig. \ref{fig:pressure}. There is assumed that the pressure build-up on the plates initiates when the piston reaches a constant threshold $z_p$.
\begin{equation}
\label{eq: zp}
p=\max(0,\frac{z-z_p}{z_M-z_p}) p_{hc}  
\end{equation}
The torque transfer due to plate pressure is modeled with parameter $\alpha$, based on the geometry of the plates and fluid characteristics. A naive approach to model the transferred clutch torque $T_c$ would be to switch between Couette flow and plate friction once $z$ equals $z_p$ in a discrete way. However, the sudden torque shift would cause unrealistic behavior. Therefore, a transient function  $\delta_t \in [0,1]$ is defined (Fig. \ref{fig:pos}) to have a smooth transition between the aforementioned sources of torque transfer.
\begin{equation*}
T_c(\omega_1,\omega_2,p,z) = \delta_t(z) \cdot \alpha p+(1-\delta_t(z) )\cdot\frac{\gamma}{z_{M}-z}\cdot(\omega_1-R \omega_2)
\end{equation*}
The expression for the counteracting clutch torque $T_c$ on the input shaft holds when $\omega_1> R\omega_2$. The driving torque of the output shaft equals $T_2=R T_c$ and is illustrated in Fig. $\ref{fig:torque}$. Hence, one can formulate the nonlinear system dynamics during the asynchronous phase by considering the torque equilibrium of both shafts.
\begin{equation}
\begin{aligned}
J_1\dot{\omega}_1&=T_1(\omega_m,\omega_1) -T_c(\omega_1,\omega_2,p,z) \\
J_2\dot{\omega}_2&=R T_c(\omega_1,\omega_2,p,z)- T_b(\omega_2)
\end{aligned}
\end{equation}.

\subsubsection{Synchronous phase}
The system shifting process arrives in the synchronous phase once an equivalent speed for both shafts is obtained ($\omega_1=R \omega_2$). Henceforth, the shafts can be considered as fully engaged and the system becomes independent of the control signal $u$. The driving torque of the output shaft equals $T_2=R T_1$ since perfect torque transmission is assumed (Fig. $\ref{fig:torque}$).
\begin{equation}
(J_1+ \frac{J_2}{R^2})\dot{\omega}_1=T_1(\omega_m,\omega_1) - \frac{1}{R}T_b\left(\frac{\omega_1}{R}\right) \\
\end{equation}

\subsubsection{Shifting time}
\begin{figure}[h!]
	\centering
	\begin{subfigure}[b]{.45\columnwidth}
		\caption{torque}
		\vspace*{-5pt}
\includegraphics[trim=6.5cm 11.7cm 7.2cm  12cm,clip=true,width=1\columnwidth]{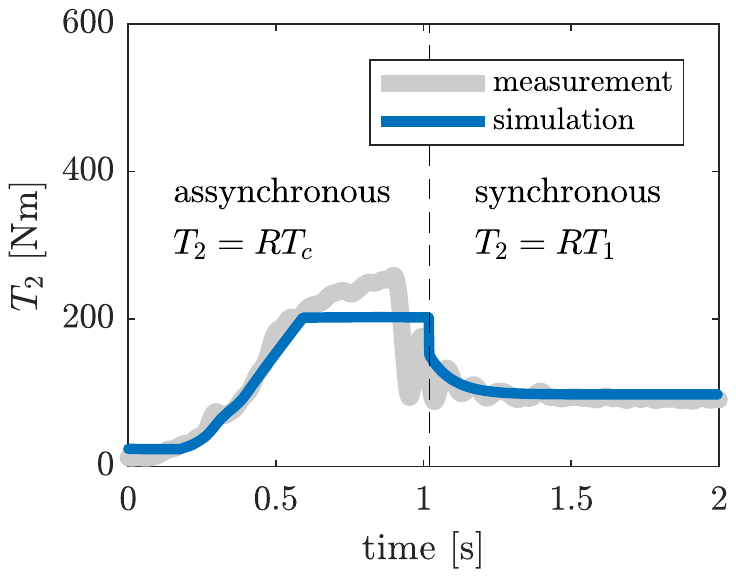}
\label{fig:torque}
	\end{subfigure}
	\begin{subfigure}[b]{.45\columnwidth}
		\caption{angular velocity }
		\vspace*{-5pt}
\includegraphics[trim=6.5cm 11.7cm 7.2cm  12cm,clip=true,width=1\columnwidth]{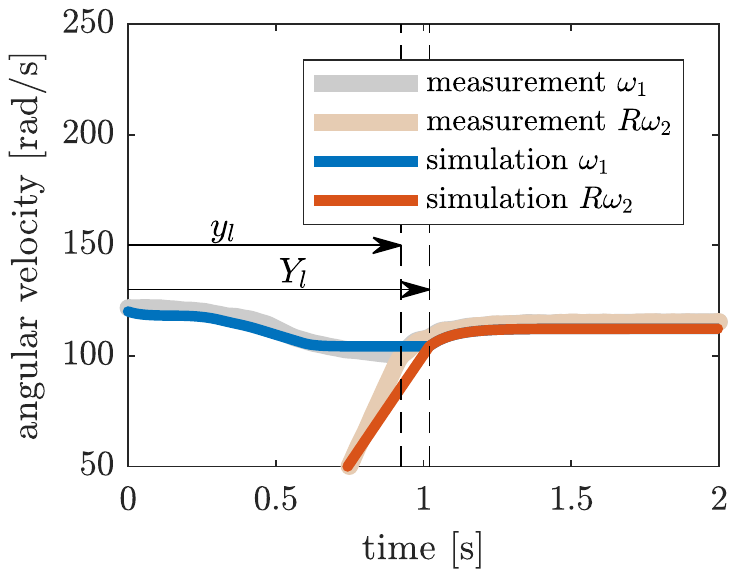}
\label{fig:omega}
	\end{subfigure}

	\caption{Comparison of dynamic simulation and signal measurements of a clutch engagement from neutral to first gear.}
	\label{fig:TS}
	\vspace{-5pt}
\end{figure}

The purpose of this dynamical model is to determine the shifting time, which is key for various applications of the wet clutch. This is determined by measuring the time interval between initialization of the feedforward control signal and the moment that both shafts obtain the same angular velocity (note that the mutual speed does not need to equal the initial input velocity). Fig. \ref{fig:omega} illustrates the acceleration of the output shaft $\omega_2$ towards a synchronous speed with the input shaft. The figure clearly illustrates the discrepancy between the measured shifting time $y_l$ and corresponding simulation $Y_l$.   

\vspace{-7.5pt}
\subsection{Practical set-up}
\begin{figure}[h]
	\centering
	\includegraphics[width=0.6\columnwidth]{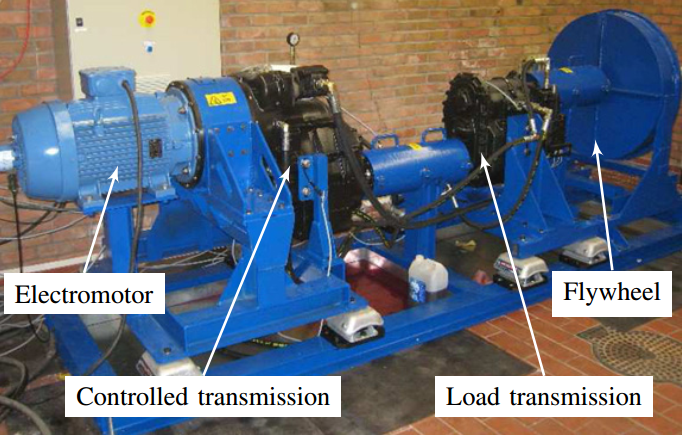}
	\caption{Wet clutch test bench} 
	\label{testbench}
	\vspace{-10pt}
\end{figure}

The test set-up considered for this paper is depicted in Fig. \ref{testbench}. An AC electric motor ($30$ \si{\kilo\watt}) is controlled to a constant speed via a high  bandwidth motor drive. The motor is connected to a controlled transmission via a torque converter. Within this transmission, the clutch can be controlled to engage with the proper gear. Furthermore, the output shaft of the clutch is connected to a load transmission. The load consists of a flywheel ($2,5$ \si{\kilogram\meter\squared}) and a brake. This system was previously used to study smooth gearbox controllers \cite{bruno2011}.
\subsection{Design of experiments}
\begin{figure}[t!]
	\centering
	\includegraphics[trim=3.7cm 12cm 4.2cm 12.4cm,clip=true,width=0.9\columnwidth]{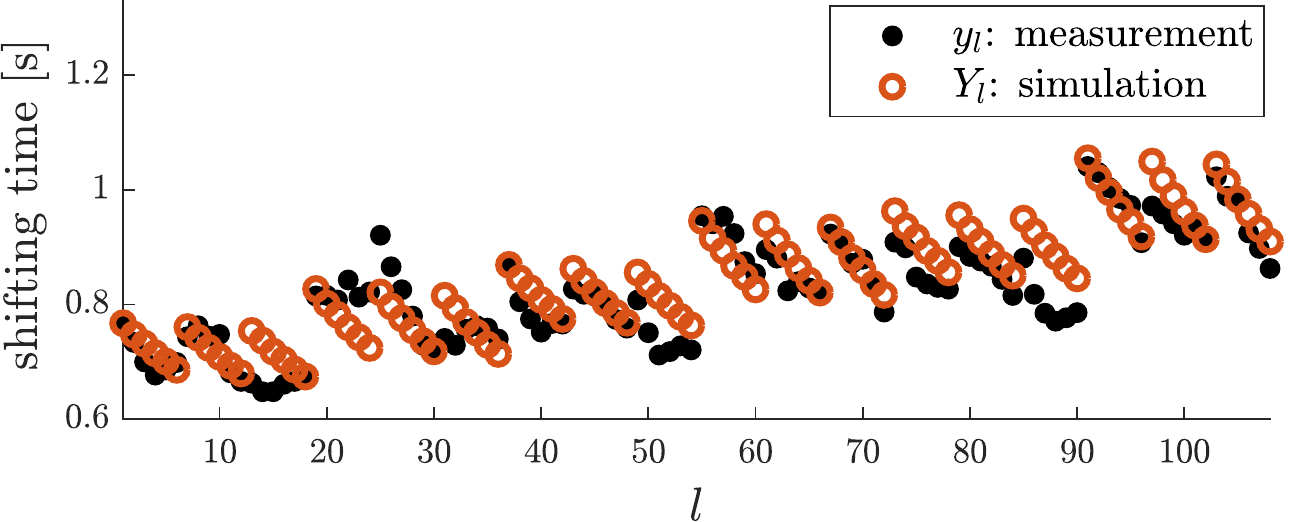}
	\caption{Experimental versus deterministic simulation results.}
	\label{fig:experiments}
	\vspace{-10pt}
\end{figure}

We experimentally tested the influence of several control and operating conditions on the shifting time $y_l$ on the wet clutch test set-up. We performed an up-shift from neutral to first gear for various control parameters and conditions, spanning a test scenario space. More specifically, 18 distinct feedforward control signals were tested. We compared 3 different values for $\Delta t$ and $u_0$ and 2 values for $\Delta u$. Further we verified 3 different motor speeds $\omega_m$ (1200, 1350 and 1500 \si{rpm}) and 2 different friction load conditions $T_b$ (low, high). For these, a full factorial design of $l\in\{1,\cdots,108\}$ experiments has been tested. As an illustration of the modeling deficiency, we compare the actual measurements, $y_l$, with the simulated shifting times, $Y_l$, in Fig. \ref{fig:experiments}. The global trend is captured by the deterministic model, however a distinct discrepancy between measurements and simulation is present. 

\vspace*{-10pt}
\subsection{Parametric uncertainty model}
\label{sec:parametric_UM}
The objective of this research is to associate and identify a probabilistic model for several of the lumped model parameters.	As mentioned in section \ref{sec:model&exp}, several of these parameters were identified empirically based on isolated subsystem measurements. The inertia of the shafts are derived from materials properties and geometrical considerations. Experiments in steady state regime are used to determine values for the viscous and Coulomb friction coefficients. On the available set-up no measurements of the oil temperature are possible and the temperature effect on the dynamics is therefore eliminated by performing all experiments after warm-up of the machine. However, our model contains various other variables that could not be determined or measured directly. Initially, we determined the parameter values using a least-squared procedure on the test scenario set. The comparison in Fig. \ref{fig:experiments} already illustrated the disadvantages of this strategy. Using a sensitivity analysis, we could identify two key parameters: $c_2$, the oil pressure bias in piston position computation (\ref{eq: pressureDyn}) and the initial piston position $z_p$ (\ref{eq: zp}). Therefore, we define the variables $x_1$ and $x_2$ that serve as scaling factor for respectively $c_2$ and $z_p$. We aim at identifying a probabilistic model for the random variable $\vectorstyle{X}=(X_1;X_2)$.
We propose a normal distribution, $\mathcal{P}_{\vectorstyle{X}}(\vectorstyle{\alpha})=\mathcal{N}(\vectorstyle{\mu},\matrixstyle{\Sigma})$, with mean $\vectorstyle{\mu}=(\mu_1;\mu_2)$ and covariance $\matrixstyle{\Sigma}$, corresponding the variable transformation, $\vectorstyle{X} = \vectorstyle{\mu} + \matrixstyle{\Sigma}\vectorstyle{\Theta},\vectorstyle{\Theta}\sim\mathcal{N}(0,\matrixstyle{I})$. Further we assume that both variables are independent so that $\matrixstyle{\Sigma} = \mathrm{diag}(\sigma_1,\sigma_2)$. Therefore, the probabilistic input model can be defined by $\vectorstyle{\alpha}=(\mu_1,\mu_2, \sigma_1,\sigma_2)$. Note that a normal transformation implicitly assumes that forward model, $\mathcal{Y}$, can be evaluated for any value $\vectorstyle{x}\in\mathbb{R}^2$. This can be avoided by clipping the distribution, constricting the stochastic domain of $\vectorstyle{X}$ to a feasible area $\vectorstyle{x} \in [\vectorstyle{x}^-, \vectorstyle{x}^+]$, i.e. we are only interested in the part of the normal distribution in between these boundaries. This is facilitated by the transformation, $\mathrm{clip}(\vectorstyle{\Theta})$. The random variables $\tilde{\vectorstyle{\Theta}}_i = \mathrm{clip}(\vectorstyle{\Theta}_i)$ should be distributed according to $\frac{1}{\Phi(\theta_i^+)-\Phi(\theta_i^-)}\varphi(\tilde{\theta}_i),\tilde{\theta}_i\in[\theta_i^-,\theta_i^+]$ where $\theta^-_i = \frac{1}{\sigma_i}(x^-_i - \mu_i)$ and $\theta^+_i = \frac{1}{\sigma_i}(x^+_i - \mu_i)$\footnote{Here $\varphi(\cdot)$ and $\Phi(\cdot)$ are standard notation for respectively the PDF and the cumulative density function (CDF) of the standard normal distribution.}. First we map $\vectorstyle{\Theta}_i$ to a uniform distribution in between the values $\Phi(\theta^-_i)$ and $\Phi(\theta^+_i)$. Then we use the inverse transform sampling method to map the uniform variable back onto the standard normal distribution but now limited to the interval $[\theta_i^-,\theta_i^+]$. In conclusion the parametric probability model is defined through the transformations
\begin{align*}
u_i &= \Phi(\theta^-_i) + \Phi(\theta_i) \left(\Phi(\theta^+_i) - \Phi(\theta^-_i)\right)\\
\tilde{{\theta}}_i &= \Phi^{-1}(u_i) \\
\vectorstyle{x} &= \vectorstyle{\mu} + \matrixstyle{\Sigma} \tilde{\vectorstyle{\theta}} = \vectorstyle{\mu} + \matrixstyle{\Sigma}\cdot \mathrm{clip}(\vectorstyle{\Theta}), ~\vectorstyle{\Theta} \sim \mathcal{N}(0,\matrixstyle{I})
\end{align*}

\begin{figure}[h]
	\centering
	\begin{subfigure}[b]{.4\columnwidth}
		\caption{model output, $l=16$}
		\vspace*{-5pt}
		\includegraphics[trim=0cm 0cm 6.4cm 0cm,clip=true,width=\columnwidth]{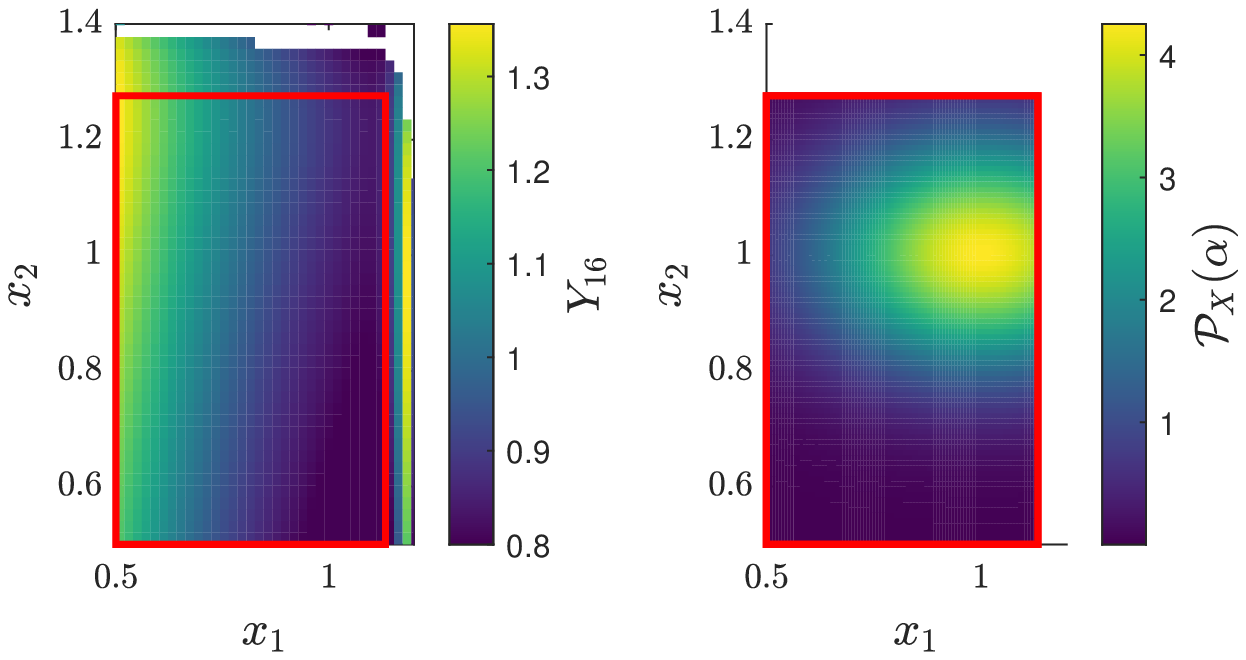}
		\label{fig:modeloutput}
	\end{subfigure}
	\begin{subfigure}[b]{.4\columnwidth}
		\caption{clipped distribution}
		\vspace*{-5pt}
		\includegraphics[trim=6.4cm 0cm 0cm 0cm,clip=true,width=\columnwidth]{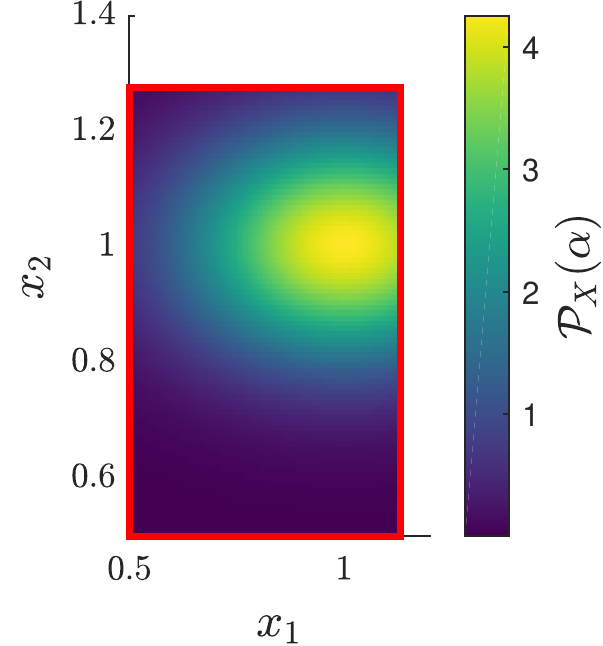}
		\label{fig:distrclip}
	\end{subfigure}
	\begin{subfigure}[b]{.9\columnwidth}
		\vspace*{-5pt}
		\caption{output distribution for each experiments}
		\vspace*{-5pt}
		\includegraphics[trim=3.5cm 12.4cm 3.8cm 12.6cm,clip=true,width=\columnwidth]{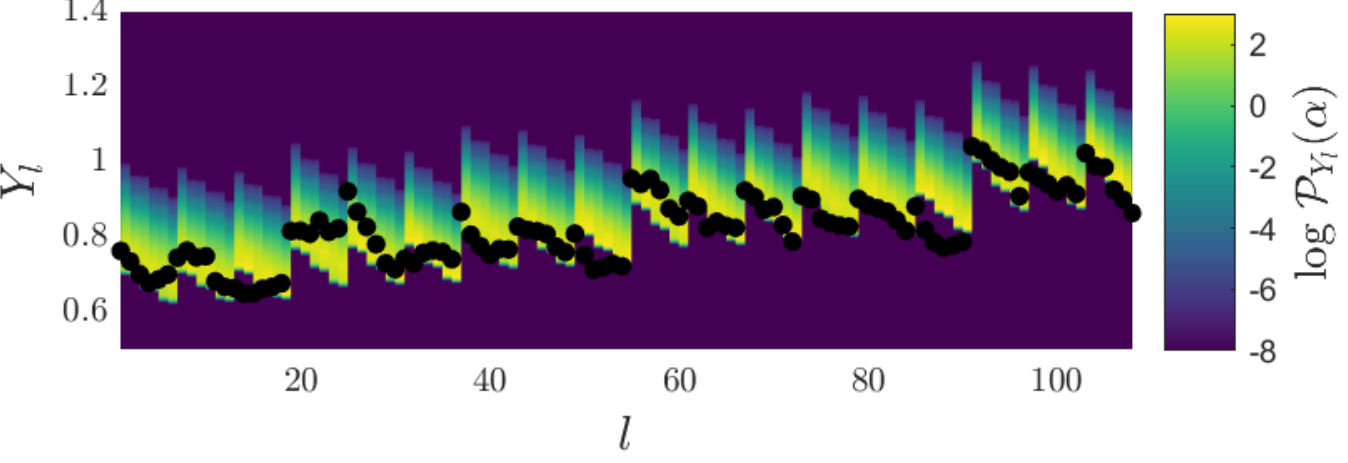}
		\label{fig:distribution_experiments}
	\end{subfigure}
	\vspace*{-10pt}
	\caption{\textit{Top left}: Simulated shifting time for $l=16$. The red box indicates the clipping domain. \textit{Top right:} Associated clipped joint distribution. \textit{Bottom}: Representation of corresponding output distributions. The $\log$-likelihood of the test scenarios for given input distributions is also given.}
\end{figure}

The clipping boundaries, $\vectorstyle{x}^-$ and $\vectorstyle{x}^+$, of the stochastic variables are chosen so that the space for which a feasible solution of the shifting time can be obtained for all 108 experiments, is maximized. Figures \ref{fig:modeloutput} and \ref{fig:distrclip} illustrate respectively the parametric model output and associated clipped normal input distribution. Figure \ref{fig:distribution_experiments} demonstrates the corresponding output distribution for each experiment. Visual comparison with Fig. \ref{fig:experiments} demonstrates that the least-squared approach treats the problem in a non appropriate manner\footnote{We may note that the least-squares approach maximizes the objective $- \sum \frac{1}{2} (y_l - \mathcal{Y}_l)^2$. Comparison with the $\log$-Likelihood approach reveals that this is equivalent with assuming that each experiment has an output probability density $\sim \exp(-\frac{1}{2}(y_l-\mathcal{Y}_l)^2)$. In other words the least-squares implicitly assumes that each experiment is normally distributed with variance $1$.}.

\vspace*{-5pt}
\section{Results and discussion}
In this section we apply the methodologies described in section \ref{sec:IUIM} on the wet clutch system as was described in section \ref{sec:application}. First we demonstrate the forward propagation of uncertainty by concatenating the techniques described in the sections \ref{sec:UP-PCE} and \ref{sec:PDF-MM}, simply to determine the corresponding output distribution by moment matching for a given input distribution. Secondly we use the MLE method to identify the optimal parameter $\vectorstyle{\alpha}^*$ that best describes observed experiments. We compare results obtained with both the Gaussian and max entropy parametric PDFs. To obtain a benchmark PDF we estimated the output distributions empirically using Monte Carlo (MC) uncertainty propagation of 200 000 simulation samples and fitted a spline curve on the associated histograms. The associated $\log L$ values are referred to as the true values.

\vspace*{-10pt}
\subsection{PDF fitting by the method of moments}
First we demonstrate the PDF estimation strategy combining high-order gPC moments with the method of moments. Corresponding the variable transformation and associated parametric probability model that were motivated in section \ref{sec:parametric_UM}, and in concordance with Table \ref{tab:wiener}, we make use of Hermite polynomials. We used polynomial order $d=4$ and corresponding univariate quadrature order $q=5$, i.e. $25$ function simulations were required to estimate the output distribution for a single experiment. Due to numerical restrictions (see section \ref{sec:algo}), we are limited to the first $m = 4$ moments. 

\begin{figure}[h]
	\centering
	\begin{subfigure}[b]{.4\columnwidth}
		\caption{probability densities}
		\vspace*{-5pt}
		\label{fig:fits_nonlin}
		\includegraphics[trim=3.8cm 9.3cm 4.1cm 9.3cm,clip=true,width=\columnwidth]{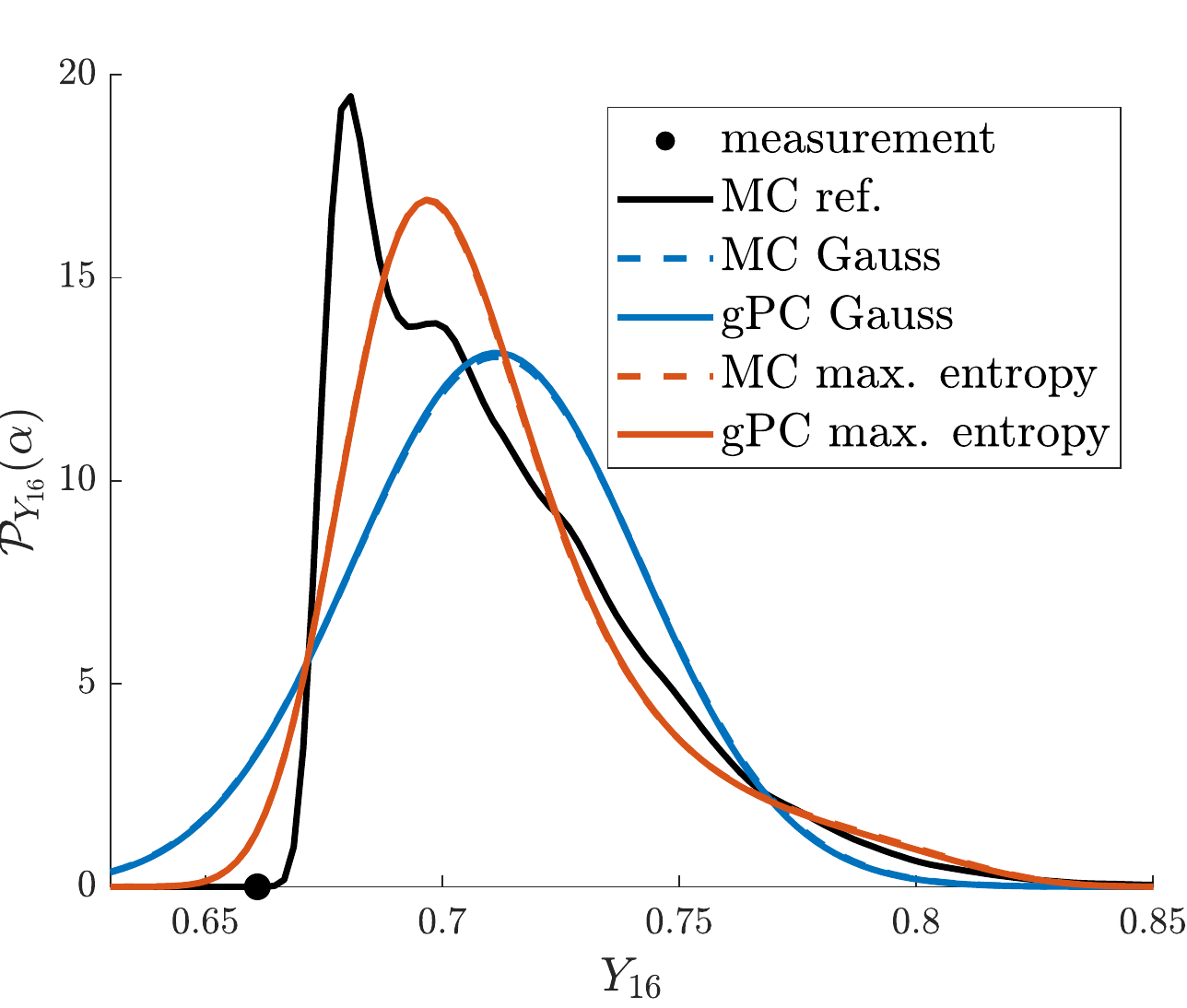}
	\end{subfigure}
	\begin{subfigure}[b]{.4\columnwidth}
		\caption{earth mover's distance}
		\vspace*{-5pt}
		\label{fig:emd_nonlin}
		\includegraphics[trim=4cm 9.5cm 4.5cm 9.6cm,clip=true,width=\columnwidth]{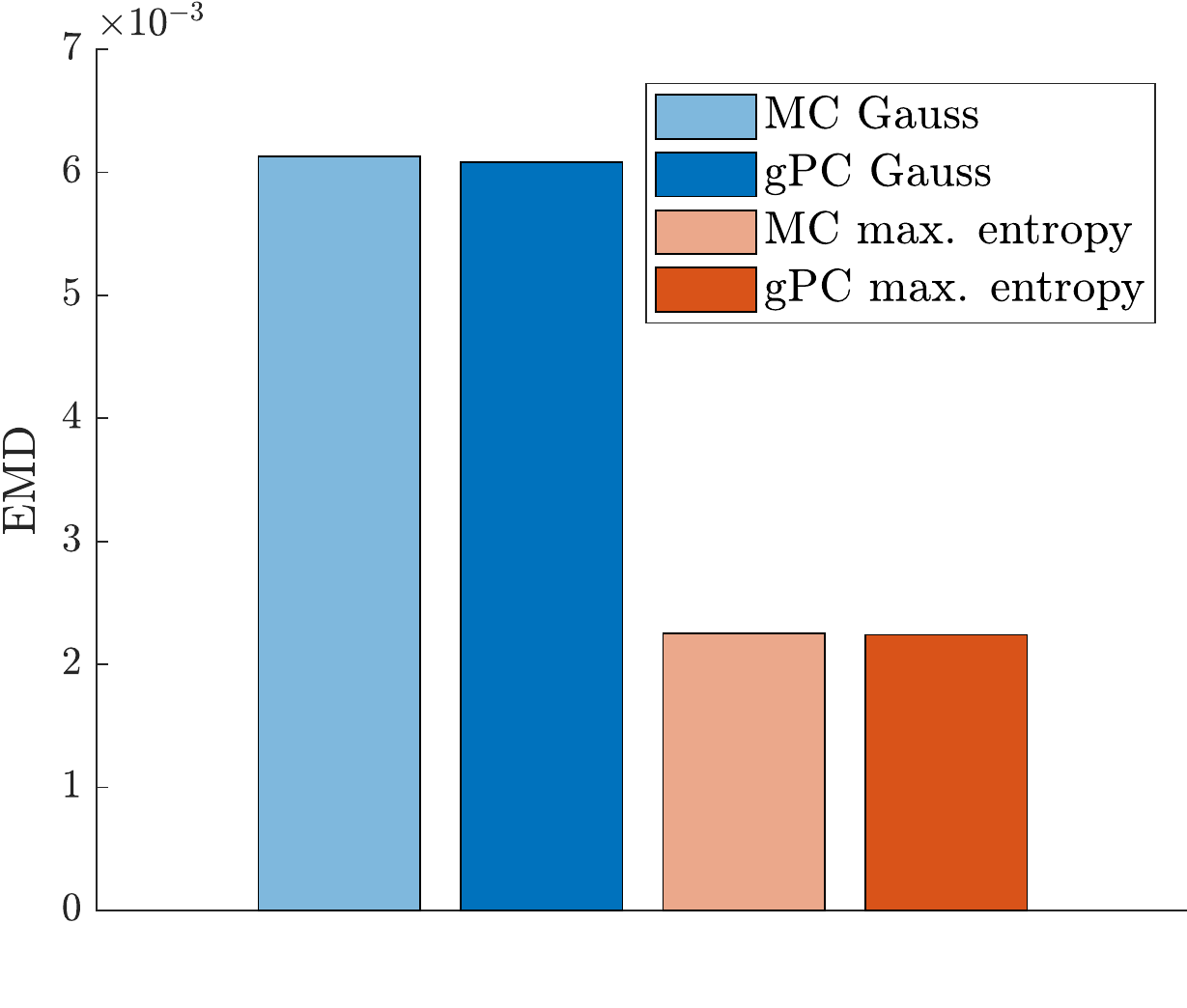}
	\end{subfigure}
	\caption{PDF estimates and corresponding EMD for given model parameters $\vectorstyle{\alpha} = (1,1,0.07,0.02)$ and experiment 16. }
	\label{fig:nonlinearfit}
	\vspace{-.75cm}
\end{figure}

Estimates for the output distribution of the 16\textsuperscript{th} experiment are visualized in Fig. \ref{fig:fits_nonlin}. We compare the MC reference distribution with Gaussian and max entropy fits. We also included PDF estimates that are based on the stochastic moments extracted from the MC sample set. Based on these results we can compare how much information could be extracted by the gPC approach with respect to the amount of information that is contained within the first four moments about the true distribution. The minor difference between the gPC PDF estimates and the MC equivalent PDF estimates suggests that gPC is a viable approach with sparse function evaluations. 

For a fair comparison besides the visual verification, we calculate the Earth Mover's Distance (EMD) (see Appendix \ref{appendix:emd}) for every estimate w.r.t. the estimated true output distribution. Different values are presented in Fig. \ref{fig:emd_nonlin}. These results confirm that high-order estimates obtained through the extended gPC framework are capable of extracting statistical information from significantly fewer function evaluations when compared to the brute force methodologies such as MC.

\subsection{Maximum log-likelihood estimation}
We demonstrated that the high-order gPC moment matching strategy provides cheap yet accurate output PDF estimates for given input uncertainty. Consequently it can be engaged as an efficient alternative to evaluate the $\log$-Likelihood of a number of experiments at a mere fraction of the computational cost associated to brute force strategies such as MC. Next we apply the method to identify the optimal parameter set $\vectorstyle{\alpha}^*$ associated to the probability model described in section \ref{sec:parametric_UM}. To solve problem (\ref{eq:MLE}), we used a genetic algorithm (GA) with population size 50 using the \Matlab ~optimization toolbox. We compared optimization results for both Gaussian and max entropy PDF abstractions. The performance of the optimal parameter sets are verified using MC estimates of the $\log$-Likelihood corresponding the identified probability models.

\subsubsection{Gaussian output distribution}
\begin{figure}[t!]
	\centering
	\begin{subfigure}[b]{.9\columnwidth}
		\caption{benchmark output distributions using MC}
		\vspace*{-3pt}
		\includegraphics[trim=3.5cm 12.4cm 3.8cm 12.6cm,clip=true,width=\columnwidth]{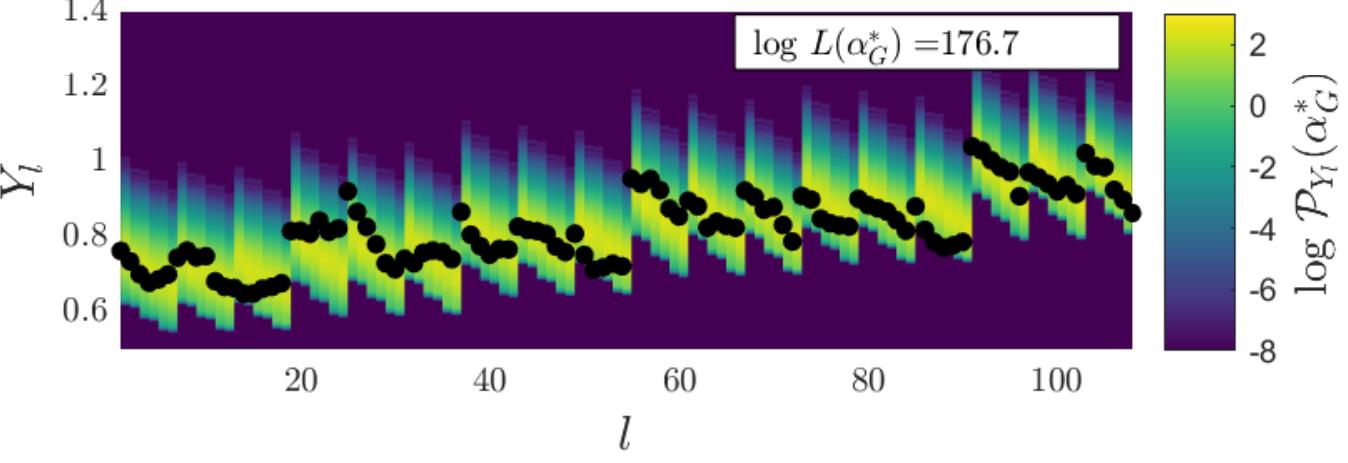}
		\label{fig:gauss1}
	\end{subfigure}
	\begin{subfigure}[b]{.9\columnwidth}
		\vspace*{-10pt}
		\caption{Gaussian distributions using gPC moment matching}
		\vspace*{-3pt}
		\includegraphics[trim=3.5cm 12.4cm 3.8cm 12.6cm,clip=true,width=\columnwidth]{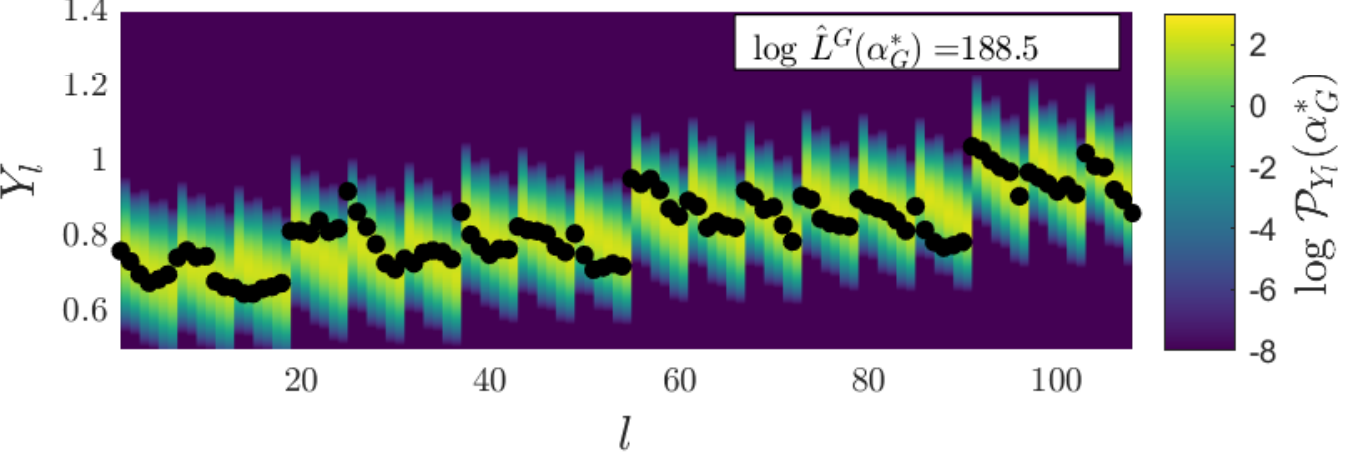}
		\label{fig:gauss2}
	\end{subfigure}
	\vspace*{-15pt}
	\caption{Representation of corresponding output distributions using the optimal parameter set, $\alpha_G^*$.}
	\label{fig:gauss}
	\vspace{-3pt}
\end{figure}

Using the Gaussian output distribution and gPC moment estimates, the following optimal parameter set was identified, $\vectorstyle{\alpha}^*_{G}=(0.91,0.55,0.077,0.014)$. Figure \ref{fig:gauss} compares measurement with the corresponding output distributions. The solution has an estimated $\log$-Likelihood of $\log\hat{L}^{G}(\vectorstyle{\alpha}^*_{G})=188.5$ using the moment matching approach. When we evaluate the $\log$-Likelihood using the MC approach, we obtain a smaller value $\log L(\vectorstyle{\alpha}^*_{G})=176.7$.

\subsubsection{Maximum entropy output distribution}
\begin{figure}[t!]
	\centering
	\begin{subfigure}[b]{.9\columnwidth}
		\caption{benchmark output distributions using MC}
		\vspace*{-3pt}
		\includegraphics[trim=3.5cm 12.4cm 3.8cm 12.6cm,clip=true,width=\columnwidth]{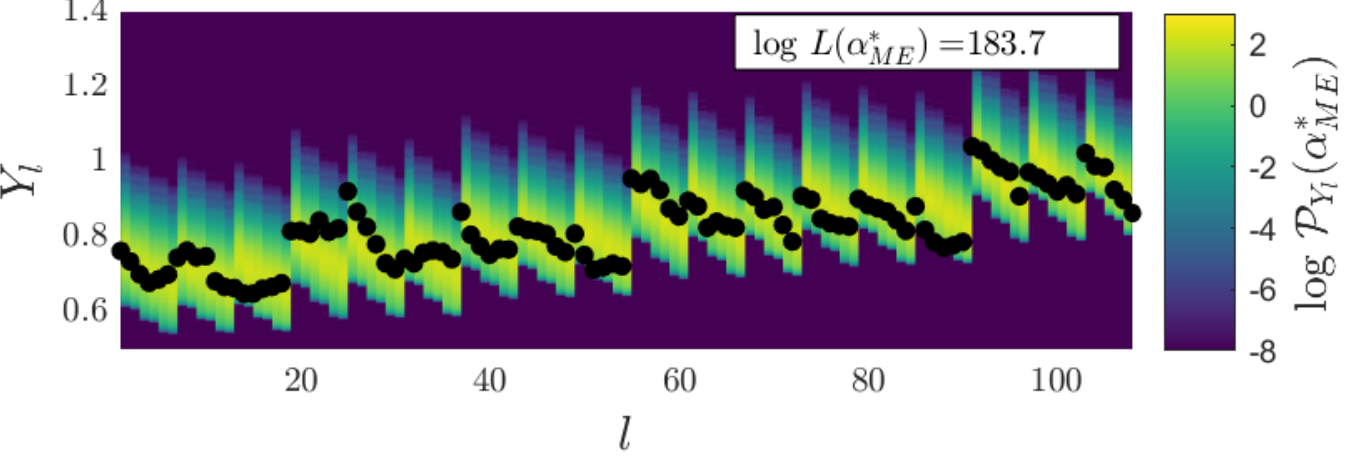}
		\label{fig:ME1}
	\end{subfigure}
	\begin{subfigure}[b]{.9\columnwidth}
		\vspace*{-10pt}
		\caption{max entropy distributions using gPC moment matching}
		\vspace*{-3pt}
		\includegraphics[trim=3.5cm 12.4cm 3.8cm 12.6cm,clip=true,width=\columnwidth]{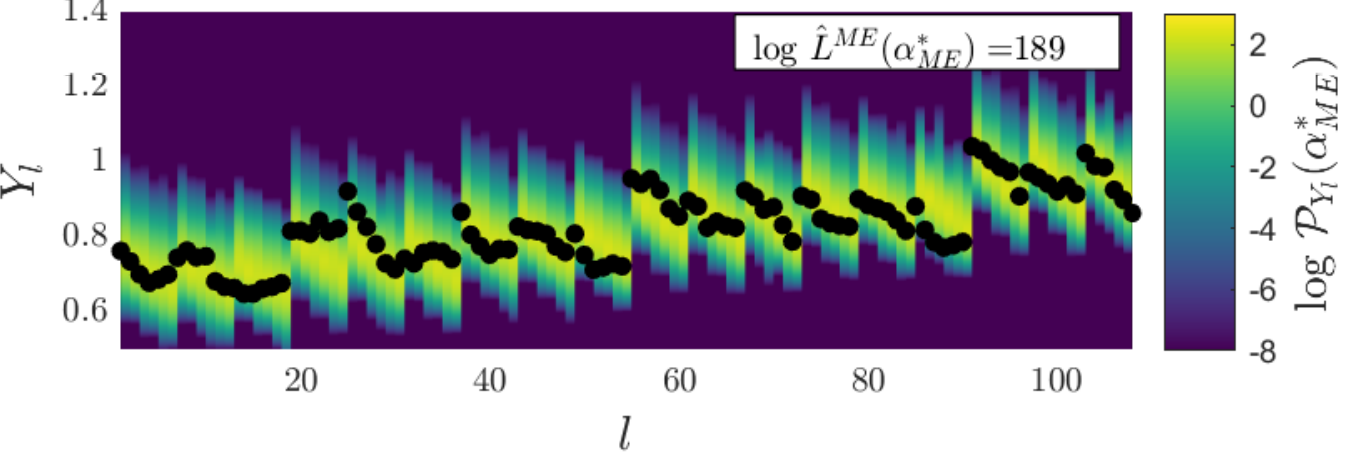}
		\label{fig:ME2}
	\end{subfigure}
	\vspace*{-15pt}
	\caption{Representation of corresponding output distributions using the optimal parameter set, $\alpha_{ME}^*$.}
	\label{fig:ME}
	\vspace{-3pt}
\end{figure}

Using the max entropy distribution we obtain the following optimal parameter, $\vectorstyle{\alpha}^*_{ME}=(0.91,0.55,0.081,0.046)$. Note that we retrieve the same mean as with the Gaussian output distributions, $\vectorstyle{\mu}$, but that both covariance values have increased. Figure \ref{fig:ME} visualizes the corresponding output distributions and measurements. Using the max entropy moment matching approach, the $\log$-likelihood is estimated on $\log \hat{L}^{ME}(\vectorstyle{\alpha}^*_{ME})=189$. Verification using MC delivers a slightly smaller $\log$-Likelihood $\log L(\vectorstyle{\alpha}^*_{ME})=183.7$. Note that the difference is significantly smaller compared to the $\log$-Likelihood mismatch obtained with the Gaussian PDF estimates.

\vspace*{-10pt}
\subsection{Discussion}
With respect to the global $\log L$ estimates, we remark the following. In both the max entropy and the Gaussian case the $\log L$ is overestimated using the proposed gPC moment matching method. However, for the max entropy distribution, the mismatch between true $\log L$ and estimated $\log L$ is significantly smaller (approximately 55\%). Moreover the true $\log L$ obtained with $\vectorstyle{\alpha}^*_{ME}$ exceeds that obtained with $\vectorstyle{\alpha}^*_{G}$ by 4\%. Both of these observations confirm that the use of high-order stochastic moments benefits the identification of an optimal input probability model and that the gPC framework can be used to obtain accurate estimates of these moments.

\begin{figure}[h!]
\centering
\includegraphics[trim=3.1cm 11.4cm 4.5cm 11.6cm,clip=true,width=.9\columnwidth]{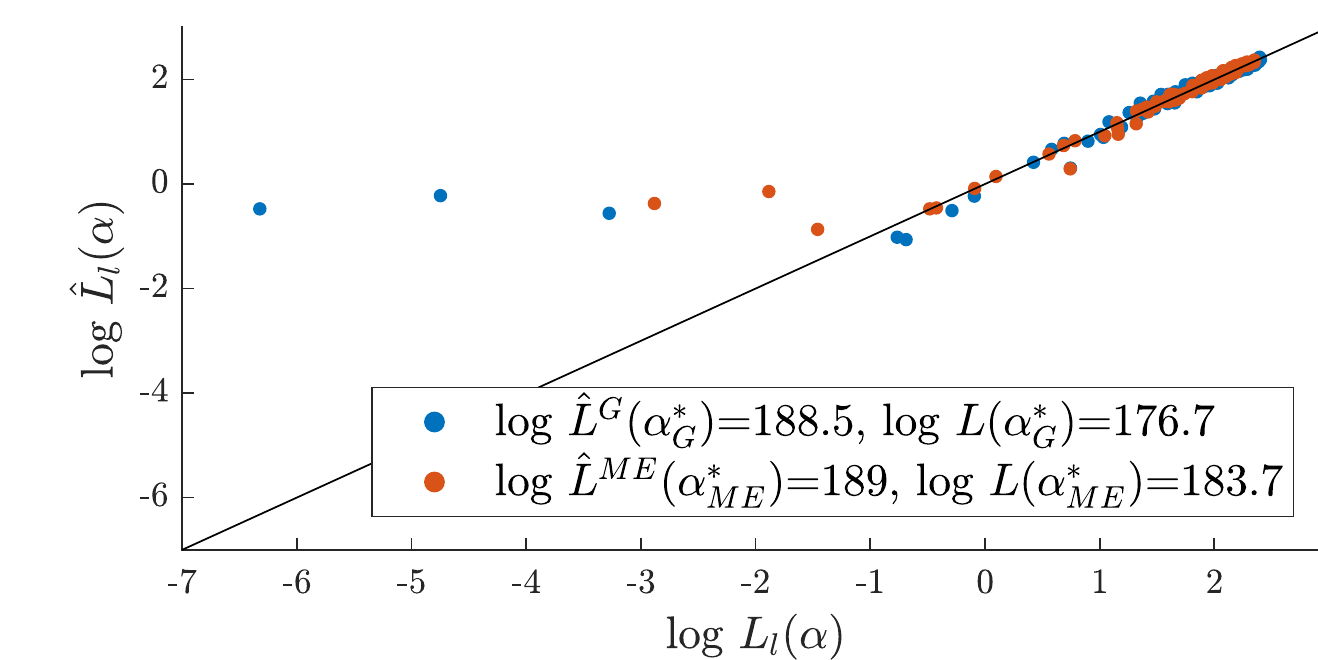}
\vspace*{-3pt}
\caption{Comparison of individual $\log$-Likelihood estimates $\hat{L}_l(\vectorstyle{\alpha})$ and corresponding reference $\log$-Likelihood $L_l(\vectorstyle{\alpha})$.}
\label{fig:LLHcomp}
\vspace{-3pt}
\end{figure}

\begin{figure}[H]
	\centering
	\includegraphics[trim=2.8cm 11.3cm 3.3cm 11.8cm,clip=true,width=1\columnwidth]{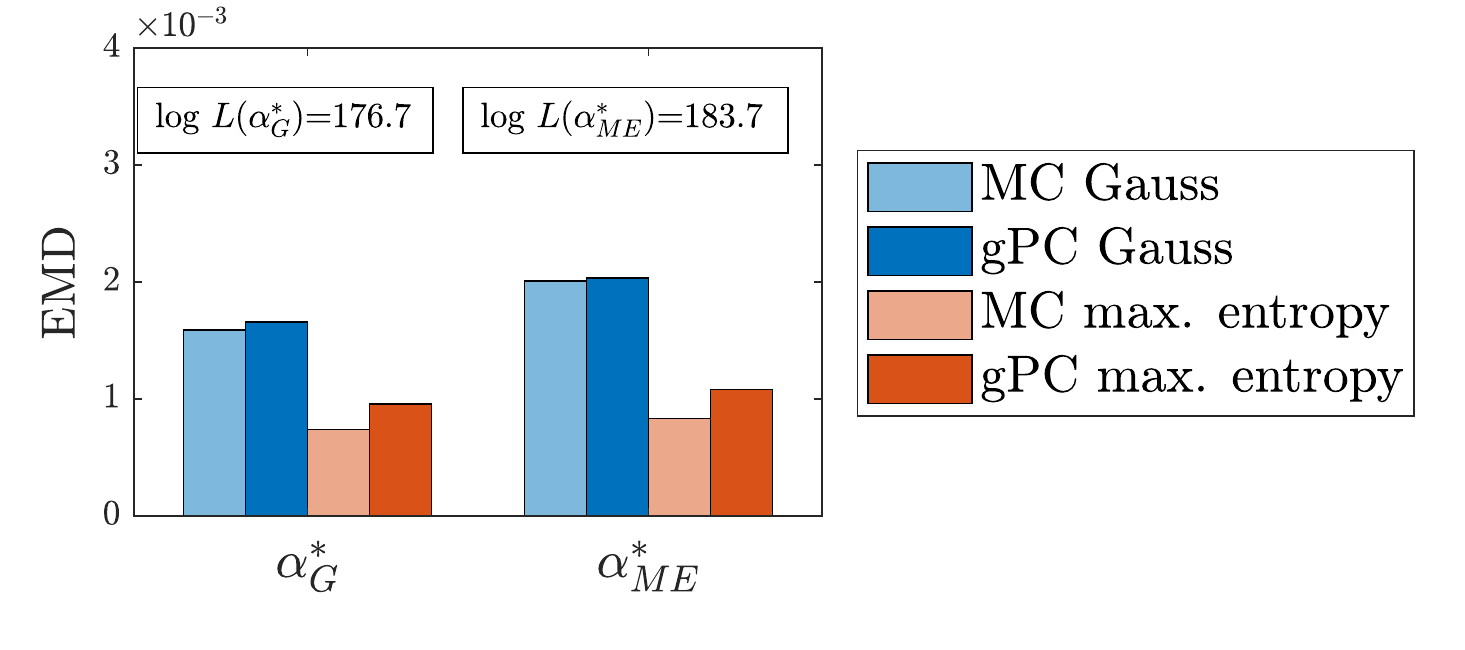}
	\caption{EMD of different PDF approximations.}
	\label{fig:rescomp}
\end{figure}

Figure \ref{fig:LLHcomp} compares the estimates $\log \hat{L}_l^{G}(\vectorstyle{\alpha}^*_{G})$ and  $\log \hat{L}_l^{ME}(\vectorstyle{\alpha}^*_{ME})$, for each individual experiment $l \in \{1,\cdots, 108\}$, w.r.t. to their true value. One may note that for the majority of the experiments the $\log L$ is estimated with great precision for either output distribution. This indicates that for the majority of the experiments, the true output distribution is about Gaussian and no additional information is contained within the high-order moments. However for a small subset of experiments, the Gaussian overestimates the true value significantly due to its incapacity of modeling asymmetric distributions. As a result, measurements that are highly improbable and would strongly affect the global $\log L$ are penalized less by the Gaussian distribution. This effect was already visualized in Fig. \ref{fig:nonlinearfit}. 
Following the argument made above, we compare the EMD metric averaged over all experiments in Fig. \ref{fig:rescomp}, as it proofs to be valuable to quantify the estimated output distribution's accuracy. We found that the use of the maximum entropy distribution with 4 moments could reduce the EMD by approximately 42\% for $\alpha_G^*$ and 47\% for $\alpha_{ME}^*$ when compared with the averaged EMD value obtained with Gaussian output distributions. The minor difference compared to the EMD of the PDF estimates based on the stochastic moments directly extracted from the MC sample set indicates that the gPC can correctly estimate the statistical moments. The improved accuracy can therefore be fully attributed to the higher order moments gPC framework.

\begin{figure}[h!]
	\centering
	\begin{subfigure}[b]{0.49\columnwidth}
		\caption{MAE for $\vectorstyle{\alpha}^*_G$}
		\label{fig: evalComp_Gauss}
		\includegraphics[trim=6cm 11.4cm 6.5cm 11.7cm,clip=true,width=1\columnwidth]{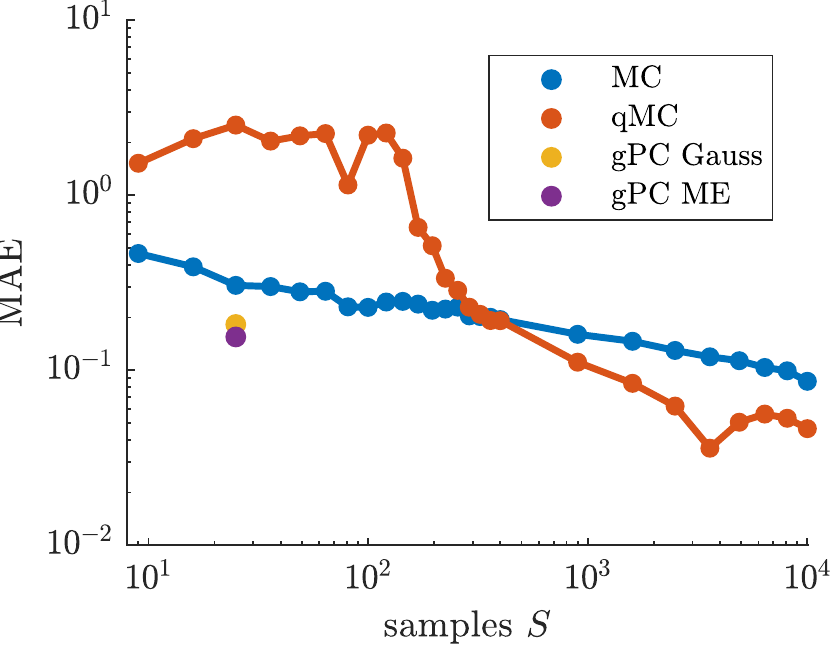}
	\end{subfigure}
	\begin{subfigure}[b]{0.49\columnwidth}
		\caption{MAE for $\vectorstyle{\alpha}^*_{ME}$}
		\label{fig: evalComp_ME}
		\includegraphics[trim=6cm 11.4cm 6.5cm 11.7cm,clip=true,width=1\columnwidth]{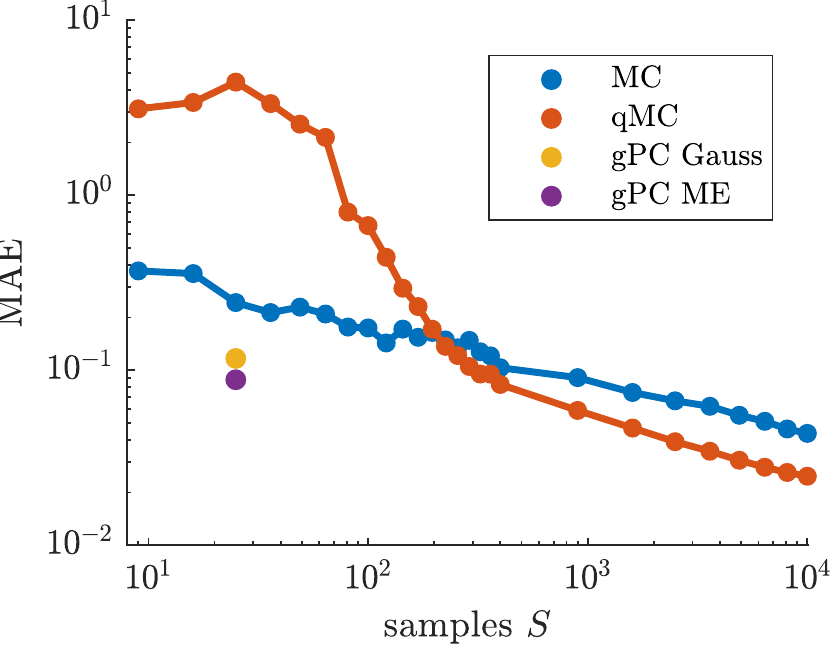}
	\end{subfigure}
	\vspace*{-3pt}
	\caption{Influence of sample density on estimation accuracy of  $\log L_l(\vectorstyle{\alpha}^*)$.}
	\label{fig:evalComp}
	\vspace{-3pt}
\end{figure}

Figure \ref{fig:evalComp} illustrates the mean absolute error (MAE) of the $\log$-Likelihood estimates of all individual experiments $l \in \{1,\cdots, 108\}$ compared to the MC reference obtained by 200 000 simulation samples.  The accuracy of the brute force propagation techniques increases with respect to the number of samples used for uncertainty propagation. The quasi Monte Carlo (qMC) technique considers an equally distributed grid of $S$ samples. The gPC results require a magnitude less expensive function evaluations to achieve the same accuracy as the brute force techniques. The interpolated values $S_{MQ}$ and $S_{qMQ}$ are depicted in Table \ref{table:overview}. A comparison between the gPC techniques indicates that the MAE could be reduced by 15.3\% for $\vectorstyle{\alpha}_G^*$ and 24.8\% for $\vectorstyle{\alpha}_{ME}^*$ due to the incorporation of higher order estimates in the novel gPC framework.

\begin{table}[h!]
	\centering
	\caption{Summary of log-Likelihood estimation.}
	\label{table:overview}
\begin{tabular}{|l|l|l|l|lll|}
	\hline
	& $\log L$	& PDF est. & MAE  & $S_{gPC}$ & $S_{MC}$ & $S_{qMC}$ \\ \hline
	\multirow{2}{*}{$\alpha^*_G$} & \multirow{2}{*}{176.7} & Gauss &0.183&25&458&577 \\
								  &								  & ME &0.155&25&629&1171 \\ \hline 
\multirow{2}{*}{$\alpha^*_{ME}$} & \multirow{2}{*}{183.7} & Gauss &0.117&25&265&370 \\
&								 & ME &0.088&25&383&1009\\ 	\hline 	
	
\end{tabular}
\end{table}

\section{Conclusion}
In this paper we proposed and discussed an efficient numerical method that is tailored to parametric model uncertainty identification using maximum likelihood estimation. Here the objective is to find a parametric input distribution that best explains a series of observations. A novel method is proposed to determine the output distribution for given input probability model making use of the generalized Polynomial Chaos (gPC) expansion framework. We extended the gPC framework so that high-order stochastic moments can be obtained efficiently. These high-order estimates can then be fed to a Gaussian or maximum entropy distribution. This strategy enables a reduction in the required number of function evaluations to obtain an adequate estimation of the statistical moments of the output distribution. The method is applied to calibrate a model for the shifting time for wet clutch engagement based on a series of measurements verified in the lab. It is shown that output distribution estimations via high-order moment matching excels traditional identification methods that are based on mean and variance estimates whilst the computational cost remains the same due to efficient techniques that were developed. The high-order moment matching resulted into a $\log$-likelihood increase of about 4\% since the accuracy of the estimated output probability density function could be improved up to 47\% compared to Gaussian distributions. This methodology reveals high potential when scaling up the number of uncertain parameters that need to be identified through inverse uncertainty identification. In addition to the inclusion of parametric uncertainty, future work will investigate the incorporation of model structure uncertainty.

\section*{Acknowledgments}

Wannes De Groote holds a doctoral grant strategic basis research  (3S07219) of the Fund for Scientific research Flanders (FWO).
Some of the computational resources and services used in this work were provided by the VSC (Flemish Supercomputer Center), funded by the Research Foundation - Flanders (FWO) and the Flemish Government – department EWI. This study was performed in the frameworks of the EVIT ICON project. This research was supported by Flanders Make, the strategic research center for the manufacturing industry.

\appendix
\section{Maximum entropy distribution}
\label{appendix:med}
The maximum entropy distribution, $\hat{p}$ provides a solution to the classical problem where an estimate of univariate density $p$ is sought from knowledge of its moments \cite{mead1984maximum}. The extent to which a density may be determined from its moments is a topic of discussion in mathematical literature. In practice only a finite number of moments, say $K+1$, are usually available so that there exists an infinite variety of functions whose first $K+1$ moments coincide and a unique reconstruction of $p$ is simply impossible. To remedy this issue, $\hat{p}$ is sought to maximize $\operatorstyle{W}[p] = -\int_{\mathcal{X}} p(x) \log p(x)\text{d}x$  under the condition that the first $K+1$ moments are equal to the true moments $\mu_k$. The Lagrangian $L$ corresponding to this problem is
\begin{equation}
\label{eq:originalL}
\operatorstyle{L}[p,\vectorstyle{\lambda}] = \operatorstyle{W}[p] + \mathsmaller{\sum}_{k=0}^K \mathsmaller{\int}_{\mathcal{X}} \lambda_k(x^k - \mu_k) p(x) \text{d}x
\end{equation}

Nullifying the functional variation to $p$ yields
\begin{equation*}
\delta_p \operatorstyle{L} = 0 \Rightarrow \hat{p}(x|\vectorstyle{\lambda}) = \exp \left(-\mathsmaller{\sum}_{k=0}^K \lambda_k x^k\right)
\end{equation*}

Without loss of generality we may further assume that $\hat{p}$ is normalized so that $\mu_0 = 1$. Accordingly, we can express $\lambda_0$ in function of the remaining Lagrangian multipliers
\begin{multline*}
\mathsmaller{\int}_{\mathcal{X}} \exp \left(-\mathsmaller{\sum}_{k=0}^K \lambda_k x^k\right) \text{d}x = 1 \\
\Rightarrow \lambda_0 = \log \mathsmaller{\int}_{\mathcal{X}} \exp \left(-\mathsmaller{\sum}_{k=1}^K \lambda_k x^k\right) \text{d}x
\end{multline*}

One may substitute this solution into the original Lagrangian (\ref{eq:originalL}) so to obtain the dual unconstrained problem which can be solved accordingly. 
\begin{equation*}
\min_{\vectorstyle{\lambda}}\Gamma(\vectorstyle{\lambda}) = \lambda_0 + \mathsmaller{\sum}_{k=1}^K \lambda_k \mu_k
\end{equation*}

\section{Earth mover's distance}
\label{appendix:emd}
The Earth Mover's distance (EMD) is a metric for the dissimilarity between two signatures that are each a compact representation of a distribution. In formal mathematics the metric is known as the Wasserstein distance between two probability distributions. The metric was coined in computer science due to an apparent analogy to its definition that can be modeled as the solution to a transportation problem \cite{rubner2000earth}.

Given are two signatures $P=\{(u_i,\vectorstyle{x}_i)\}_{i=1}^n$ and $Q=\{(v_j,\vectorstyle{y}_j)\}_{j=1}^m$. An insightful interpretation is that $u_i$ and $v_j$ represent the amount of dirt at the respective positions $\vectorstyle{x}_i$ and $\vectorstyle{y}_j$. The EMD between the signatures $P$ and $Q$ is defined as the minimal (normalized) work required to reconfigure $P$ into $Q$ moving around the dirt. Formally that is
\begin{equation*}
\begin{aligned}
\mathrm{EMD}(P,Q)=~ &\min_{F=\{f_{ij}\}} &&\frac{\sumnolim_{ij} f_{ij}d(\vectorstyle{x}_i,\vectorstyle{y}_j)}{\sumnolim_{ij} f_{ij}} \\
&\text{subject to}&& \mathsmaller{\sumnolim}_i f_{ij} \leq v_j \\
&&&\mathsmaller{\sumnolim}_j f_{ij} \leq u_i \\
&&&\mathsmaller{\sumnolim}_{ij} f_{ij} = \min\left\{\mathsmaller{\sumnolim}_i u_i ,\mathsmaller{\sumnolim}_j v_j \right\} \\
&&&f_{ij} \geq 0
\end{aligned}
\end{equation*}
where $f_{ij}$ represents the dirt transferred from position $\vectorstyle{x}_i$ to $\vectorstyle{y}_j$, and, $d(\vectorstyle{x}_i,\vectorstyle{y}_j)$ the distance to be covered for that transport.

The transportation problem above is a special linear programming problem. 

\section{Derivation of equation (\ref{eq:moments2.0})}
\label{appendix:mathderiv}

Details are given about the derivation of formula (\ref{eq:moments2.0}). We initiate the derivation from equation (\ref{eq:moments}). First we expand the power of the sum using the multinomial theorem. Then we regroup the terms so that the high-order inner product is isolated from the coefficients. Finally we substitute expression $\prod_{j=1}^{n} \phi^{(j)}_{\uline{k}(j)}$ for $\psi_{\uline{k}}$ so to obtain equation (\ref{eq:moments2.0}).

\begin{equation}
\begin{aligned}
\mu^{(d)}_m &= \int_{\mathcal{X}} \left(\sum_{i\in\mathcal{I}} c_i \psi_i\right)^m f_{\vectorstyle{X}} \text{d}\vectorstyle{x} \\
&= \int_{\mathcal{X}} \sum_{\uline{i}\in\mathcal{I}(m,p)} \binom{m}{\uline{i}} \prod_{k=1}^{p} c_k^{i_k} \psi_k^{i_k} f_{\vectorstyle{X}} \text{d}\vectorstyle{x} \\
&= \sum_{\uline{i}\in\mathcal{I}(m,p)} \binom{m}{\uline{i}} \cdot  \left\langle \prod_{k=1}^{p} \psi_k^{i_k} \right\rangle \cdot \prod_{k=1}^{p}c_k^{i_k} \\
&= \sum_{\uline{i}\in\mathcal{I}(m,p)} \binom{m}{\uline{i}} \cdot  \prod_{j=1}^{n} \left\langle \prod_{k=1}^{p} {\phi^{(j)}_{\uline{k}(j)}}^{i_k} \right\rangle \cdot \prod_{k=1}^{p}c_k^{i_k} 
\end{aligned}
\end{equation}

\bibliographystyle{model1-num-names}
\bibliography{references}

\end{document}